\documentclass[MSNbibl,nameyear,seceqn,dvips]{arxstspdf}
\usepackage{flushend}
\usepackage{stfloats}
\usepackage{graphicx}

\volume{28}
\issue{4}
\pubyear{2013}
\firstpage{564}
\lastpage{585}
\doi{10.1214/13-STS445} 

\makeatletter
\setattribute{abstract}    {skip} {20\p@}
\setattribute{keyword}     {skip} {8\p@}

\newcommand{\eqref}[1]{(\ref{#1})}
\renewcommand{\citep}[1]{(\citeauthor{#1}, \citeyear{#1})}
\newcommand{\Pn}{$P_n$}
\newcommand{\argmin}{\mathop{\mathrm{arg\, min}}}

\makeatother

\begin{document}
\begin{frontmatter}

\title{Wind Energy: Forecasting Challenges for Its Operational
Management}
\runtitle{Wind Energy: Forecasting Challenges}

\begin{aug}
\author{\fnms{Pierre} \snm{Pinson}\corref{}\ead[label=e1]{ppin@dtu.dk}} %
\runauthor{P. Pinson}
\affiliation{Technical University of Denmark}
\address{Pierre Pinson is Professor in the Modeling of Electricity
Markets,
Centre for Electric Power and Energy,
Department of Electrical Engineering,
Technical University of Denmark,
Elektrovej 325(058),
2800 Kgs. Lyngby, Denmark \printead{e1}.}

\end{aug}

%
\begin{abstract}
Renewable energy sources, especially wind energy, are to play a larger
role in providing electricity to industrial and domestic consumers.
This is already the case today for a number of European countries,
closely followed by the US and high growth countries, for example,
Brazil, India and China. There exist a number of technological,
environmental and political challenges linked to supplementing existing
electricity generation capacities with wind energy. Here,
mathematicians and statisticians could make a substantial contribution
at the interface of meteorology and decision-making, in connection with
the generation of forecasts tailored to the various operational
decision problems involved. Indeed, while wind energy may be seen as an
environmentally friendly source of energy, full benefits from its usage
can only be obtained if one is able to accommodate its variability and
limited predictability. Based on a short presentation of its physical
basics, the importance of considering wind power generation as a
stochastic process is motivated. After describing representative
operational decision-making problems for both market participants and
system operators, it is underlined that forecasts should be issued in a
probabilistic framework. Even though, eventually, the forecaster may
only communicate single-valued predictions. The existing approaches to
wind power forecasting are subsequently described, with focus on
single-valued predictions, predictive marginal densities and space--time
trajectories. Upcoming challenges related to generating improved and
new types of forecasts, as well as their verification and value to
forecast users, are finally discussed.
\end{abstract}

%
\pdfkeywords{Decision-making, electricity markets, forecast verification, Gaussian copula, linear and nonlinear
regression, quantile regression, power systems operations, parametric and nonparametric predictive densities,
renewable energy, space-time trajectories, stochastic optimization}

\end{frontmatter}

\section{Introduction}\label{sec1}

Increased concerns related to climate evolution and energetic
independence have supported the necessary technological and regulatory
developments to broaden the energy mix all around the world, with a
particular emphasis placed on renewable energy sources \citep
{Letcher08}. Among the various candidates, wind energy showed the most
rapid and consistent deployment of power generating capacities. By June
2012, the cumulative installed wind power capacity worldwide had
reached 254 GW, and it is still increasing at a rapid pace [\cite
{WWEA12}]. Besides all the mathematical and statistical challenges in
the development of turbines (aerodynamics, materials and structures,
etc.), and in the deployment of wind energy capacities (e.g., wind
resource estimation, logistics optimization), those relating to power
systems operations and electricity markets have attracted substantial
and growing interest over the last 3 dec\-ades. This is since, in
contrast with conventional generation means, wind power generation
cannot be scheduled at will, except maybe by curtailing the output of
the wind turbines. Wind power is produced as the wind blows: the
dynamics of wind power generation are the result of a nonlinear
conversion and filtering of wind dynamics through the turbines' rotor
and electric generator. It makes that the traditional operational
methods used for conventional generators cannot directly apply to wind
energy. For that reason, of the various renewable energy sources, wind,
solar, wave and tidal energy are often referred to as stochastic power
generation, owing to their inherent variability and uncertainty.

Wind energy is by far the renewable energy source that has attracted
the most attention of researchers and practitioners. It is clear,
however, that a number of operational and economic issues will be the
same for the other forms of renewable energy sources. In practice, such
challenges require the modeling and forecasting of the wind power
generation process at various temporal and spatial scales, to be
subsequently used as input to decision-making. Our objective here is to
give an overview of these forecasts and of challenges stemming from
their generation and verification. It is to be noted that forecasting
is only one aspect of better accommodating renewable energy generation,
such as that from the wind into existing power systems and electricity
markets. For instance, from a more general perspective of investment,
regulation and policy, even the way wind energy should be compared to
conventional technologies challenges traditional practice \citep
{Joskow11}. Similarly, when assessing resource adequacy (i.e., making
sure that the overall generating capacity is sufficient to meet demand
at all times) and competition in electricity markets, it is argued that
the impact of renewable energy sources on market dynamics ought to be
accounted for \citep{Wolak13}.

The most classical statistical problem involving wind energy is that of
resource assessment, that is, focusing on unconditional distributions
of wind speed and the corresponding potential power generation. In
practice, this is based on estimating margin\-al wind distributions given
a (potentially limited) sample of wind measurements on site and/or in
the vicinity of the sites of interest. Even though these marginal
distributions are highly valuable for the optimal siting and design of
wind farms, they have nearly no value for the operational management of
wind power generation: they give an unconditional picture only, hence,
they do not give information on the volatile and conditional
characteristics of wind and power dynamics at the relevant spatial and
temporal scales. A succession of two papers published in the \textit
{Journal of Applied Meteorology} in 1984 is a symbol of the transition
from models for limiting distributions only to dynamic models. There,
the seminal work of \citet{Conradsen84} on fitting Weibull
distributions to samples of wind speed measurements of various lengths
is literally followed by that of \citet{Brown84}, which certainly was
the first paper looking at dynamic (linear time-series) models for the
prediction of wind speed and corresponding power generation. Not so
long after, \citet{Haslett89} bridged the gap between the two by
focusing on the dynamic spatio-temporal structure of wind speed over
Ireland and its implications for the wind energy resource. Since then,
ample research was performed on stochastic dynamic models for the
prediction of wind power generation at lead times between a few minutes
and up to several days ahead, accounting or not for spatial effects.
For an exhaustive review of the state of the art in that field, the
reader is referred to \citet{Giebel11}, while a solid introduction to
the physical concepts involved can be found in \citet{Lange06}. Our
state of knowledge today is that optimal decision-making involving wind
power generation calls for predictions generated in a probabilistic
framework. These should inform of uncertainties through predictive
marginal densities, but also potentially of spatio-temporal
dependencies through trajectories, which are known as scenarios in the
operations research literature. As a very recent example of how
forecasts in their most simple deterministic form, or as space--time
trajectories, may be used as input to operational problems, the reader
is referred to \citet{Papavasiliou13}, focusing on a unit commitment
problem (i.e., the least-cost dispatch of available generation units)
under transmission network constraints.

Wind power generation is first introduced in Section~\ref{sec:wpstochproc} as a stochastic process observed at discrete points
in space and in time. Subsequently, in order to underline the
importance of probabilistic forecasts (in contrast to deterministic,
single-valued forecasts), Section~\ref{sec:opchal} describes
representative decision problems involving wind energy in power systems
operations and its participation in liberalized electricity markets.
Section~\ref{sec:mod} then covers the various types of forecasts used
today and to be employed in the future for optimal decision-making. The
paper ends in Section~\ref{sec:concl} with a discussion that covers
{(i)} the current and foreseen challenges for forecast improvement,
{(ii)} the proposal of thorough and appropriate verification
frameworks, and {(iii)} the importance of bridging the gap between
forecast quality and value.

\section{Wind Power Generation as a Stochastic Process}
\label{sec:wpstochproc}

Some of the early works on dynamic modeling and forecasting of wind
power generation were cast in a physical deterministic framework, as,
for instance, \citet{Landberg94} on local wind conditions, and
similarly for the follow-up study (Landberg, \citeyear{Landberg99}) on power
generation. Today however, there is a broad consensus that wind power
generation should be modeled as a stochastic process, whatever the
spatial and temporal scales involved. A~part of uncertainty comes from
our lack of knowledge of all the physical processes involved, combined
to our limited ability to account for all of them in mathematical and
statistical models. There may also be some inherent uncertainty in the
data generating process. The choice for appropriate distributions may
not be straightforward.

The physical basics of wind power generation are presented in
Section~\ref{ssec:phys}. Definitions and notation are introduced
subsequently in Section~\ref{ssec:notat}. Finally, the Western Denmark
data set is described in Section~\ref{ssec:wden}. It will be used for
illustrating the different forms of forecasts that will be described
throughout the paper.

\subsection{Physical Basics of Wind Power Generation}
\label{ssec:phys}

The generation of electric power from the wind relies on atmospheric
processes. The power output of a single wind turbine is a direct
function of the strength of the wind over the rotor swept area.\break
Coarsely simplifying the meteorological aspects involved, winds
originate from the movement of air masses from high to low pressure
areas: the larger the difference in pressure, the stronger the
resulting winds. On top of that come the boundary layer effects,
complexifying wind behavior due to natural obstacles, friction
effects, the nature of the surface itself, temperature gradients, etc.
The boundary layer is formally defined as the lower part of the
atmosphere where wind speed is affected by the surface. The resulting
level of complexity makes that the characteristic features of wind
variability may be better described in the frequency domain
(Mur~Amada and Bayod~R{\'u}jula, \citeyear{MurAmada10}). Our state of the knowledge on wind dynamics in the
boundary layer, and, more generally, mesoscale meteorology, is today
still limited: resulting models of wind characteristics have systematic
and random errors.

Wind speed exhibits fluctuations over a wide range of frequencies.
Those in the order of days are induced by the movement of synoptic
weather patterns, that is, by general changes in weather situations.
These are modeled within global weather models such as those run at the
European Centre for Medium-range Weather Forecasts (ECMWF, in the UK)
and at the National Centers for Environmental Prediction (NCEP, in the
US), among others. Those models encompass well-known dynamics of state
variables for the global weather, while wind components are a
by-product derived from the evolution of these state variables. In
terms of forecasting, several directions are thought of today for
improving the estimation of the initial state of the Atmosphere and
also to better account for potential uncertainties in the model and its
parametrization \citep{Palmer12}.

Fluctuations referred to as diurnal and semi-diurnal cycles (with
periods of 24 and 12 hours) are mainly induced by thermal exchanges
between the surface (land or sea) and the Atmosphere. Their magnitude
varies as a function of local climate and seasons. At these time
scales, the phenomena involved are fairly well known, though certain
aspects like their impact on wind profiles (i.e., the way wind
evolves with height) still are a subject of active research, for
example, \citet{Pena10}. At frequencies in the order of minutes to
hours, local effects potentially including the presence of cumulus
clouds, convective cells, precipitation, waves (for offshore sites),
etc. are the drivers of wind speed variations. Here, the physical and
mathematical aspects may become more challenging owing to the
combination of a substantial number of interacting physical processes.
Higher frequencies (seconds to a few minutes, not considered in the
present paper) see a dominance of turbulence effects, which are a
particular concern for the structural design of turbines, fatigues
studies and, potentially, control. Finally, at the other end of the
spectrum, very low frequencies also seen as long-term wind trends, have
attracted increased attention recently since human activity and climate
evolution may potentially impact surface winds at these time scales;
see \citet{Vautard10}, for instance. In the following, emphasis is
placed on time scales in the order of minutes to days, where existing
meteorological challenges include the better understanding of the
physical processes and their interaction, as well as their modeling.

%
\begin{figure}

\includegraphics{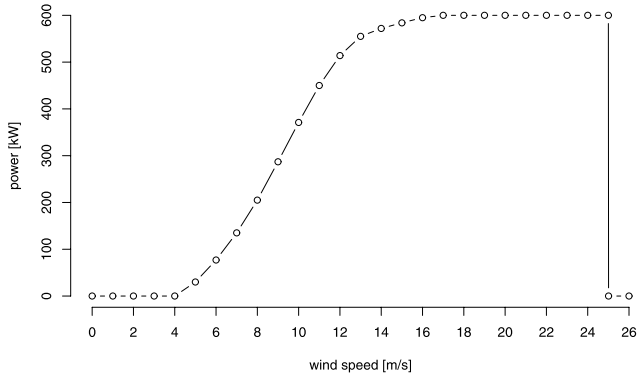}

\caption{Power curve of the Vestas V44 (600~kW) wind turbines installed at the Klim
wind farm, for an air density of 1.225 kg/m$^3$.}
\label{fig:tpc} \vspace*{-3pt}%
\end{figure}

Wind speed is the meteorological variable of most relevance to power
generation. The process of the conversion of wind to electric power for
a single wind turbine is described by its power curve. It is also
influenced by air density (being a function of temperature, pressure
and humidity) to a minor extent. Power curves for different turbines
roughly have the same shape for all manufacturers and turbine types. In
order to discuss and illustrate what manufacturer's (i.e., theoretical)
and observed power curves may look like, let us take the example of the
Klim wind farm in Western Denmark. It is composed of 35 Vestas V44 wind
turbines having a capacity of 600 kW each, yielding a nominal capacity
of 21 MW. The nominal capacity of a wind turbine or of a wind farm is
the power output it generates within the range of wind conditions over
which it was designed to operate, ideally. Figure~\ref{fig:tpc} depicts
the power curve for a V44 turbine. The power production is null below
the cut-in wind speed (4 m/s), then sharply augments between the cut-in
and rated wind speeds (16~m/s). At rated speed, it reaches its nominal
power \Pn. The power production is nearly constant between rated and
cut-off wind speeds (here 25 m/s). At cut-off speed, the turbine stops
for security reasons. This power curve example is for a fairly old wind
turbine model, since this wind farm started operating in 1996. Various
technological improvements have been permitted to lower cut-in and
rated wind speeds, which are today between 2 and 4 m/s for the former
one and between 12 and 15 m/s for the latter one. Moreover, cut-off
wind speeds may reach up to 34 m/s. In a general manner there may also
be a difference between the maximum (peak) and nominal power values (up
to 10--20\%). Most importantly, the nominal capacity of today's wind
turbines is up to 7--8 MW.

A power curve such as in Figure~\ref{fig:tpc} is a theoretical one,
since it gives the power output of a single turbine exposed to ideal
wind conditions as if in a wind tunnel (i.e., not altered by
obstacles, without turbulence and for the turbine always perfectly
facing the wind), for a given air density. In practice, however, wind
turbines are almost always gathered in wind farms with potentially a
mix of different turbine types. The combination of these individual
power curves will not be the same as that of any of the individual
turbine types. Besides, depending upon the prevailing wind direction,
some of the turbines within a wind farm may mask the others---the
so-called shadowing effect, therefore reducing the wind seen by these
turbines. This combines with additional surrounding topographic and
orographic effects (i.e., hills, forest, etc.), making that the various
turbines within a wind farm are constantly seeing different wind
conditions, which also are different from the free-stream wind at a
reasonable distance away from the wind farm. Consequently, the
resulting wind farm power curve has features far more complex than the
theoretical power curves provided by the manufacturers for individual
wind turbines.

\begin{figure}

\includegraphics{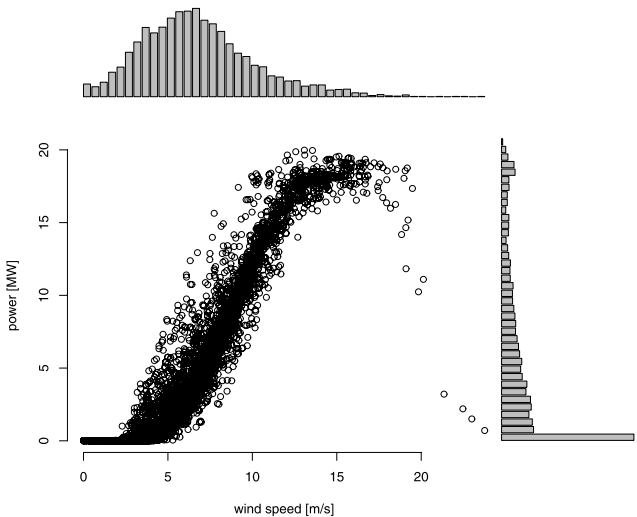}

\caption{Example empirical power curve for the Klim wind farm over a 6-month period in
2002, based on hourly measurements of wind speed and corresponding
power output. Marginal distributions of wind speed and power are also
represented above, and, respectively, right of, the power curve itself.}\label{fig:pc}
\end{figure}

Figure~\ref{fig:pc} depicts the empirical power curve of the Klim wind
farm based on hourly wind speed (at 10~m above ground level) and power
measurements collected over the first 6 months of 2002. For both types
of measurements, a record for a given point in time corresponds to the
average value over the preceding hour. Measurement errors in power and
wind speed observations certainly contribute to the scatter of data
observed. However, the main reason for that scatter is the impact of
other meteorological variables, as, for example, wind direction and air
density, on the power generation from the wind farm. For measured wind
speeds of 5 m/s the observed power output of the wind farm varies
between 0 and 7 MW, while for wind speeds of 10 m/s, that same power
output may be between 6 and 15 MW. Other reasons for these variations
include natural changes in the environment of the wind farm, aging of
turbine components, etc. At the turbine, wind farm or portfolio (i.e.,
a group of geographically distributed wind farms, though jointly
operated) level, all empirical power curves exhibit characteristics
differing from those of theoretical ones, also with a substantial
scatter of observations. Other interesting empirical power curves for
wind farms in Crete, as well as challenges related to their modeling,
were recently discussed by \citet{Jeon12}.

%
%

\subsection{Preliminaries and Definitions}
\label{ssec:notat}

Owing to the combination of complex physical processes, and since we
may not have a perfect understanding of all these processes anyway, it
is acknowledged that one should account for a random uncertainty
component in the modeling of energy generation from wind turbines. Wind
power is therefore considered as a discrete stochastic process, that
is, as a set of random variables $Y_{s,t}$ observed at discrete points
in time~$t$ and in space $s$. Depending upon the practical setup, it
may reduce to a temporal process with a set of random variables $Y_t$
for successive times, for instance, if concentrating on a single wind
farm or on a fixed (geographically spread) portfolio, or to a spatial
process with a set of random variables $Y_s$ for a given time but for
various locations, for instance, if looking at maps of wind energy
resource over a region. The corresponding realizations of the process
are denoted by $y_{s,t}$ in the more general spatio-temporal case, or,
more simply, by $y_t$ and $y_s$ in the temporal and spatial cases,
respectively. The notation $f$ and $F$ are used for probability
density and cumulative distribution functions (abbreviated p.d.f. and c.d.f.)
of the random variables involved, with appropriate indices.

Wind power generation as a stochastic process exhibits features that
can be seen as fairly unique, even though relevant parallels with
stochastic processes for other renewable energy sources, in meteorology
and hydrology or in economics and finance, exist. Some of these
characteristic features come from the very nature of wind, while some
others are directly linked to the process of converting the energy in
the wind to electric power. First of all, wind components and resulting
wind speed have a combination of dynamic and seasonal features, which
may vary depending on local wind climates and regions of the world.
Besides, when focusing on spatial and temporal scales of relevance to
power systems operations and electricity markets, the various
meteorological phenomena involved induce switches in the dynamic
behavior of wind fluctuations and in their predictability, yielding a
nonstationary process [see the discussion by \citet{Vincent10}, for
instance]. Inspired by models developed in the econometrics literature,
the existence of successive periods with different levels of
predictability of wind speeds was first captured with a Generalised
Auto Regressive Conditional Heteroscedastic (GARCH) model by \citet
{Tol97}, though focusing on coarser daily wind records.

In parallel, the conversion of the energy in the wind to electric power
acts as a nonlinear transfer function (as represented in Figure~\ref{fig:pc}) making wind power generation a nonlinear and bounded
stochastic process. There may even be smooth temporal changes in this
nonlinear transfer function owing to, for example, aging of equipment,
changes in external environment, etc. The transfer function shapes the
predictability of wind power generation. Consequently, conditional
densities of wind power generation should be seen as non-Gaussian, with
their moments of order greater than 1 directly influenced by their mean
(\citeauthor{Lange05}, \citeyear{Lange05}; Bludszuweit, Dom{\'i}nguez-Navarro and
Llombart, \citeyear{Bludszuweit08}).
Truncated Gaussian, Censored Gaussian
and Generalized Logit--Normal distributions were proposed as relevant
candidates for the modeling of conditional densities of wind power
generation \citep{Pinson12}. In terms of stochastic differential
equations, this would translate to having a state-dependent diffusion
component. The flat parts of the transfer function also yield
concentration of probability mass at the boundaries, potentially
requiring to consider wind power generation as a discrete-continuous
mixture, similar to precipitation, for instance.

After proposing a suitable model structure, and estimating its
parameters, such a model may be employed to simulate time-series of
wind power generation for one or several locations, for instance, as
input to power systems and market-related analysis. In most cases,
however, forecasting is the final application. Predictions fed into
operational decision problems always are for future points in time and
rarely for new locations at which no observations are available.
Consequently, even though spatial aspects are of crucial interest, the
problem at hand is mainly seen as a temporal forecasting problem. The
set of $m$ locations is denoted by
%
%
%
\begin{equation}
\mathbf{s} = \{s_1,s_2,\ldots,s_m\}.
\label{eq:slocs}
\end{equation}
In parallel, the set of $n$ lead times is
%
%
%
\begin{equation}
t+\mathbf{k} = \{t+1, t+2, \ldots, t+n\}, \label{eq:kltimes}
\end{equation}
where $n$ is the forecast length. Lead times are spaced regularly and
with a temporal resolution equal to the sampling time of the process
observations. Since the power generation process is bounded, it can be
marginally normalized, so that
%
%
%
\begin{equation}
\mathbf{y}_{\mathbf{s},t+\mathbf{k}} \in[0,1]^{mn}. \label{eq:normy}
\end{equation}

At time $t$ the aim is to predict some of the characteristics of
%
%
%
\begin{eqnarray}\label{eq:multprocsk}
&&\mathbf{Y}_{\mathbf{s},t+\mathbf{k}}
\nonumber
\\[-8pt]
\\[-8pt]
\nonumber
&&\quad = \{ Y_{s,t+k}; s= s_1,
\ldots,s_m , k = 1, \ldots, n \},
\end{eqnarray}
a multivariate random variable of dimension $m \times n$ in the
complete spatio-temporal case, or of
%
%
%
\begin{equation}
\mathbf{Y}_{t+\mathbf{k}} = \{ Y_{t+k}; k = 1, \ldots, n \},
\label{eq:multprock}
\end{equation}
a multivariate random variable of dimension $n$, in the simpler setup
where spatial considerations are disregarded.

In the most general case, the forecaster issues at time $t$ for the set
of lead times $t+\mathbf{k}$ a probabilistic forecast $\hat
{F}_{\mathbf
{s}, t+\mathbf{k}|t}$ (here a predictive c.d.f.) describing as faithfully
as possible the c.d.f. $F_{\mathbf{s}, t+\mathbf{k}}$ of the random
variable $\mathbf{Y}_{\mathbf{s}, t+\mathbf{k}}$, given the information
available up to time $t$. It hence translates to a full description of
marginal densities for every location and lead time, as well as
spatio-temporal dependencies among the set of $m$ locations and $n$
lead times. This clearly comprises a difficult problem, both in terms
of generating such forecasts and also for their verification.
Consequently, since degenerate versions of that problem may be more
tractable, a~number of them have been dealt with in the literature, for
instance, for the forecasting of marginal densities for each location
and lead time individually, or even by forecasting some summary
statistics (more precisely, mean and quantiles) of these marginal
densities only.

The combination of all uncertainties, related to physical aspects to be
accounted for in the models, but also in connection with the
data-generating process, obviously is going to impact the quality of
the resulting forecasts. In Section~\ref{sec:mod} some of the most
common approaches to forecasting will be reviewed. They all tend to
disregard the specific contributions of physical and data-generating
processes to forecast quality. Alternative proposals in a robust
forecasting framework could therefore be beneficial.

\subsection{The Western Denmark Data Set}
\label{ssec:wden}

A data set for the Western Denmark area is used as a basis for
illustration. It consists of wind power measurements as collected by
Energinet.dk, the transmission system operator in Denmark. This region
has one of the highest wind power penetrations (i.e., the share of
wind power in meeting the electric energy demand) in the world,
consistently between 25 and 30\% over the last few years.

%
\begin{figure*}
\begin{tabular}{@{}c@{\hspace*{2.5cm}}c@{}}

\includegraphics{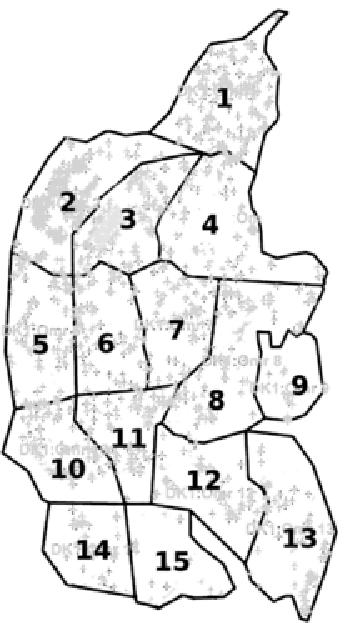}

&
\begin{tabular}{@{}lcc@{}}
\hline
\textbf{Agg. zone} & \textbf{Orig. zones} & \textbf{\% of capacity}\\
\hline
1 & 1, 2, 3 & 31 \\
2 & 5, 6, 7 & 18\\
3 & 4, 8, 9 & 17\\
4 & 10, 11, 14, 15 & 23\\
5 & 12, 13 & 10\\
\hline\vspace*{140pt}
\end{tabular}\vspace*{-110pt}
\end{tabular}
\caption{The Western Denmark data set: original locations for which
measurements are available, 15 control zones defined by Energinet.dk,
as well as the 5 aggregated zones. The total nominal capacity for
Western Denmark was 2.5 GW over the period covered by this data set.}
\label{fig:dk1-15zones}
\end{figure*}


%

Wind power measurements are originally available at more than 400
geographically distributed grid-connection points. Observations have an
hourly resolution over a period between 1 January 2006 and 24 October
2007. They represent average hourly power values. For operational
purposes, these are gathered in 15 so-called control zones depicted in
Figure~\ref{fig:dk1-15zones} along with their identification numbers.
The total nominal capacity slightly evolved during this period though
generally being around 2.5 GW. In order to additionally simplify this
case-study, the original 15 control zones are aggregated into 5 zones
only (see Figure~\ref{fig:dk1-15zones}), each having a different share
of the overall wind power capacity for that region. All power
measurements are normalized by the respective nominal capacities of the
5 aggregated zones. This aggregation is for the sake of example only
and could be seen as wind power generation portfolios operated by a set
of power producers in that region. Working at such a coarse spatial
resolution certainly is sufficient for some decision problems, also
simplifying modeling and estimation challenges. However, it may be that
for some applications the statistician and forecaster has to work with
the original 400-location data set, so that he has to finely analyze
and model the observed spatio-temporal dynamics; see \citet{Girard13},
for instance. This would be the case if all the owners/operators of
these individual wind farms ask for predictions in order to design
market offering strategies or for the network operator to perform very
detailed system simulations based on the impact of spatially
distributed wind power generation.

\begin{figure*}

\includegraphics{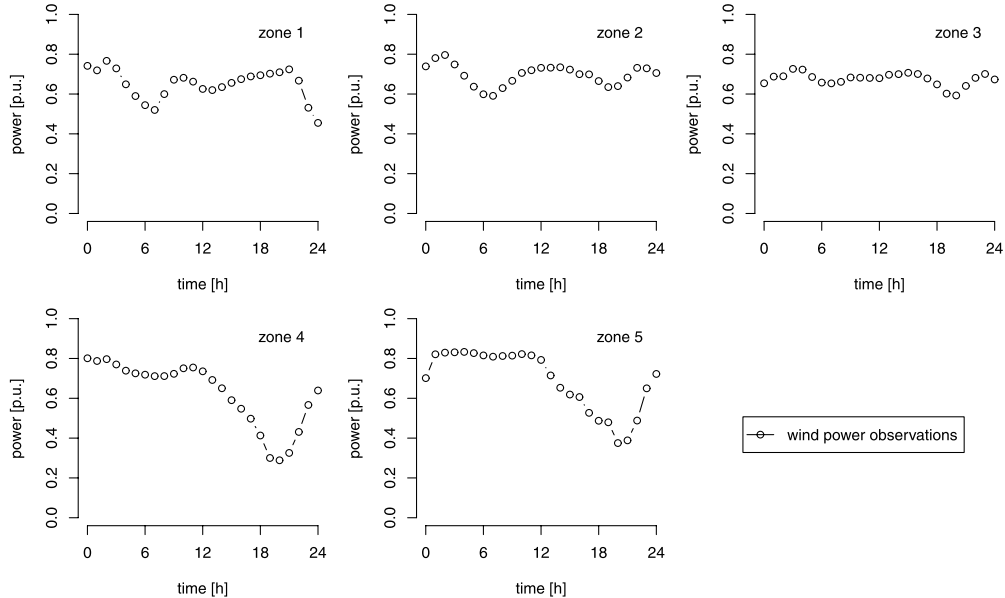}

\caption{Example episode with normalized wind power
measurements for the 5 zones of the Western Denmark data set over 24
hours, starting from the 18 March 2007 at 12 UTC.}\label{fig:mts} %
\end{figure*}

Some of the features of this data at such temporal and spatial scales
can be observed from the example episode with 24 hours of hourly wind
power measurements in Figure~\ref{fig:mts}, for the 5 aggregated zones
of Western Denmark. The spatio-temporal interdependence structure of
the wind power generation process, as induced by the inertia in weather
phenomena and resulting winds, especially results in smooth temporal
variations at each zone, individually, as well as in similarities in
the patterns observed at the various zones. These spatio-temporal
dependencies are necessarily strengthened by the aggregation procedure
employed. For instance, the drop in power generation observed in zone 4
on 19 March 2007 at 8:00 UTC (i.e., the 20th time step)
is also visible for zone 5, at the same time and with a similar
magnitude, while it may potentially be related to a drop of lesser
magnitude observed in zones 2 and 3 around the same time. Note that UTC
(for Coordinated Universal Time) is the most common standard for
referring to time in the meteorological and wind energy communities.

%

\section{Some Representative Operational Decision-Making Problems
Involving Wind Energy}
\label{sec:opchal}

Some of the representative operational decision problems are described
here, while a more extensive overview of such problems may be found in
\citet{Ackermann12}. The side of power producers is taken first, by
considering the issue of designing optimal offering strategies in
electricity markets. Subsequently taking the side of the system
operator instead (like Energinet.dk, the transmission system operator
for Western Denmark), an issue of rising importance is that of
quantifying the necessary back\-up generation to accommodate variability
and limited predictability of wind power generation. These two
decision-making problems are somehow interrelated, since the
quantification of necessary backup capacities is performed in a dynamic
way, conditional on the clearing of the electricity market. For both
types of problems, forecasts for other quantities than wind power
generation may be necessary, like load and prices. There exists
substantial literature on the statistical modeling and forecasting of
these market variables. The interested reader is referred to \citet
{Weron06} for an overview.

\subsection{Participation of Wind Energy in Electricity Markets}
\label{ssec:bid}

In a number of countries with significant wind power generation,
electricity markets are organized as electricity pools, gathering
production and consumption offers in order to dynamically find the
quantities and prices for electricity generation and consumption that
permit to maximize social welfare. These electricity pools typically
have two major stages which are the day-ahead and the balancing
markets. The electricity pool for Scandinavia, used as an example here,
is commonly known as the Nord Pool. For an overview of some the
European electricity markets and of the way they deal with wind power
generation, see \citet{Morthorst03}. A parallel overview for the case
of US electricity markets can be found in \citet{Botterud10}.

Electricity exchanges first take place in the day-ahead market for the
next delivery period, that is, the next day. Production offers and
consumption bids are to be placed for every time unit before gate
closure, occurring 12 hours before delivery in the Nord Pool, where
market time units are hourly. At the time $t$ of gate closure, wind
power producers shall propose energy offers based on forecasts with
lead times $t+k$, $k \in\{13,14,\ldots,37\}$. The market clearing is
there to match production offers and consumption bids through a single
auction process, yielding the system price $\pi^c_{t+k}$ and the
program of the market participants, that is, a set of energy blocks
$y^c_{t+k}$ to be delivered by wind power producers,\footnote{Note that
the notation $y^c_{t+k}$ is used abusively for simplification. This is
since the energy block for hour $t+k$ is necessarily equal to the
average power production value $y^c_{t+k}$ over that one-hour period.}
for every market time unit. The superscript $c$ indicates that this
combination of energy quantity and price defines a contract. Power
producers are financially responsible for any deviation from this
contract. Indeed, in a second stage, the balancing market managed by
the system operator ensures the real-time balance between generation
and load, while translating to financial penalties for those who
deviate from their contracted schedule. The prices for buying and
selling on the balancing market are denoted by $\pi_{t+k}^{b}$ and
$\pi
_{t+k}^{s}$, respectively. They are generally less advantageous than
those in the day-ahead market, fairly volatile and substantially
different from one another in a two-price settlement system like that
of the Nord Pool. The combination of the inherent uncertainty of wind
power predictions and of the asymmetry of balancing prices encourages
market participants to be more strategic when designing offering
strategies \citep{Skytte99}.

Simplifying the decision problem for clarity, potential dependencies
among time units and in space throughout the network are disregarded. A
wind power producer is seen as participating with a global portfolio of
wind power generation in the electricity market. The overall market
revenue $R_{t+k}$ is a random variable, which, given the decision
variable $y_{t+k}^c$ and the random variable $Y_{t+k}$, can be defined as
%
%
%
\begin{equation}
R_{t+k} = s_{t+k} \bigl(y_{t+k}^c \bigr)
+ B_{t+k} \bigl(Y_{t+k},y_{t+k}^c \bigr),
\label{eq:income}
\end{equation}
where the first part corresponds to the revenue from the day-ahead
market, $s_{t+k} (y_{t+k}^c) = \pi^c_{t+k} y_{t+k}^c$, while the second
is that from the balancing market, to be detailed below. Following
\citet{Pinson07} (among others), this revenue can be reformulated as a
combination of revenues and costs in a way that the decision variable
appears in the balancing market term only
%
%
%
\begin{equation}
R_{t+k} = \tilde{S}_{t+k} (Y_{t+k}) -
\tilde{B}_{t+k} \bigl(Y_{t+k},y_{t+k}^c
\bigr), \label{eq:incomer}
\end{equation}
that is, as the sum of a stochastic, though fatal since out of the
control of the decision-maker, component $\tilde{S}_{t+k}$ from selling
of the energy actually produced through the day-ahead market, minus
another stochastic component $\tilde{B}_{t+k}$, whose characteristics
may be altered through the choice of a contract $y^c_{t+k}$. The
imbalance is also a random variable, given by $Y_{t+k} - y_{t+k}^c$,
yielding the following definition for~$\tilde{B}_{t+k}$:
%
%
%
\begin{eqnarray} \label{eq:imbpricing-alt}
\qquad&&\tilde{B}_{t+k} \bigl(Y_{t+k},y_{t+k}^c
\bigr)
\nonumber
\\[-8pt]
\\[-8pt]
\nonumber
&&\quad= \cases{ %
\pi_{t+k}^{\downarrow}
\bigl(Y_{t+k} - y_{t+k}^c \bigr), &
$Y_{t+k} - y_{t+k}^c \geq0,$\vspace*{2pt}\cr
- \pi_{t+k}^{\uparrow}
\bigl(Y_{t+k} - y_{t+k}^c \bigr) , &
$Y_{t+k} - y_{t+k}^c < 0,$}
\end{eqnarray}
where $\pi_{t+k}^{\downarrow}$ and $\pi_{t+k}^{\uparrow}$ are referred
to as the regulation unit costs for downward and upward balancing,
respectively. They are readily given by
%
%
%
\begin{eqnarray}
\label{eq:def-pi+} \pi_{t+k}^{\downarrow} & = & \pi_{t+k}^{c}
- \pi_{t+k}^{s},
\\
\pi_{t+k}^{\uparrow} & = & \pi_{t+k}^{b} -
\pi_{t+k}^{c}. \label{eq:def-pi-}
\end{eqnarray}

For most electricity markets regulation unit costs are always positive,
making $\tilde{B}_{t+k} \geq0$, while the overall market revenue has
an upper bound obtained when placing an offer corresponding to a
perfect point prediction, $y_{t+k}^c=\hat{y}_{t+k|t}=y_{t+k}$. As this
is not realistic, and accounting for the uncertainty in wind power
forecasts, optimal offering strategies are to be derived in a
stochastic optimization framework. Assuming that the wind power
producer is rational, his objective is to maximize the expected value
of his revenue for every single market time unit, since this permits to
maximize revenues in the long run. Additionally considering the market
participant as a price-taker (i.e., not influencing the market
outcome by his own decision), and having access to forecasts of the
regulation unit costs ($\hat{\pi}_{t+k|t}^\downarrow$ and $\hat{\pi
}_{t+k|t}^\uparrow$), the optimal production offer $y^*_{t+k}$ at the
day-ahead market is given by
%
%
%
\begin{equation}
y^*_{t+k} = \argmin_{y^c_{t+k}} \mathbb{E} \bigl[ {B}_{t+k}
\bigl(Y_{t+k},y^c_{t+k} \bigr) \bigr].
\label{eq:opt_bid}
\end{equation}
This stochastic optimization problem has a closed-form solution, as
first described by \citet{Bremnes04}, that is, for any market time unit
$t+k$, the optimal wind power production offer is given by
%
%
%
\begin{equation}
y_{t+k}^* = \hat{F}{}^{-1}_{t+k|t}  \biggl(
\frac{\hat{\pi}_{t+k|t}^\downarrow}{\hat{\pi}_{t+k|t}^\downarrow+
\hat{\pi}_{t+k|t}^\uparrow} \biggr), \label{eq:EUM-bid}
\end{equation}
where $\hat{F}_{t+k|t}$ is the predictive c.d.f. issued at time $t$ (the
decision instant) for time $t+k$. In other words, the optimal offer
corresponds to a specific quantile of predictive densities, whose
nominal level $\alpha$ is a direct function of the predicted regulation
unit costs for this market time unit. That problem is a variant of the
well-known linear terminal loss problem, also called the newsvendor
problem \citep{Raiffa61}. It was recently revisited by \citet
{Gneiting10}, who showed that for a more general class of cost
functions [i.e., generalizing that in~\eqref{eq:imbpricing-alt}]
optimal point forecasts are quantiles of predictive densities with
nominal levels readily determined from the utility function itself,
analytically or numerically. Note that appropriate forecasts of
regulation unit costs are also needed here. It was shown by \citet
{Zugno13b} and the references therein that these may be obtained from
variants of exponential smoothing (in its basic form or as a
conditional parametric generalization) and then directly embedded in
offering strategies such as those given by~\eqref{eq:EUM-bid}.

In their simplest form, market participation problems involving wind
energy rely on a family of piecewise linear and convex loss functions,
for which optimal offering strategies are obtained in a straightforward
manner, as in the above. These only require quantile forecasts for a
given nominal level or maybe predictive densities of wind power
generation for each lead time individually. However, when complexifying
the decision problem by adding dependencies in space (e.g., spatial
correlation in wind power generation, network considerations) and in
time (e.g., accounting for the temporal structure of forecast
errors), it then requires a full description of $\mathbf{Y}_{\mathbf
{s},t+\mathbf{k}}$ (ideally in the form of trajectories), instead of
marginal densities for the whole portfolio and for each lead time
individually. The same goes for alternative strategies of the
decision-makers, for instance, if one aims to account for risk
aversion. The resulting mathematical problems do not rely on studying
specific families of cost functions, but instead translate to
formulating large scenario-based optimization problems, in a classical
operations research framework. Some of the resulting stochastic
optimization problems may be found in \citet{Conejo10}. The price-taker
assumption is also to be relaxed to a more general stochastic
optimization framework, where wind-market dependencies are to be
described and accounted for \citep{Zugno13}.

\subsection{Quantification of Necessary Power Systems' Reserves}
\label{ssec:reserves}

On the other side, the electric network operator has the responsibility
to ensure a constant match of electricity generation and consumption,
outside of the market framework described before. It involves the
quantification of so-called reserve capacities, prior to actual
operations, to be readily available if needed. This may be either for
supplementing generation lacking in the system, for example, in case of
asset outages, general loss of production and unforeseen increase in
electricity demand, or, alternatively, for lowering the overall level
of generation in the system when demand is less than production. For an
overview, see~\citeauthor{Doherty05}\break  (\citeyear{Doherty05}).

For simplicity and clarity, the timeline here is the same as for the
market participation problem described earlier. Potential dependencies
among time units and in space throughout the network (as induced by
potential network congestion) are disregarded. The system operator has
to make a decision at time $t$ (market gate closure) for all time units
$t+k$ of the following day. Reserves are to be quantified as two
numbers $q^\downarrow_{t+k}$ and $q^\uparrow_{t+k}$ for the whole power
system, for downward (when consumption is less than production) and
upward (conversely) balancing, respectively. The choice of optimal
reserve levels is linked to a random variable $O_{t+k}$ describing all
potential deviations from the chosen dispatch (consisting in the
reference values for generation and consumption at every time $t+k$).
This random variable is commonly referred to as the system generation margin.

$O_{t+k}$ can be defined as a sum of random variables representing all
uncertainties involved. These include {(i)} potential forecast
errors $\varepsilon_L$ for the electric load, {(ii)} the probability
of generation loss\break  through asset outages (assets being conventional
generators, transmission lines and other equipment), and {(iii)}
potential forecast errors $\varepsilon_Y$ for the various forms of
stochastic power generation. For simplicity, we only consider wind
power here, corresponding to the operational situation where, as in
most countries, wind power is the prominent form of stochastic power
generation. In a more general setup the combination of uncertainties
with, for example, solar and wave energy, should also be accounted for.
These various uncertainties are fully characterized by probabilistic
forecasts available at time $t$: $\hat{f}^{\varepsilon_L}_{t+k|t}$ for the
load, $\hat{f}^{G}_{t+k|t}$ for generation losses, and $\hat
{f}^{\varepsilon_Y}_{t+k|t}$ for wind power generation. This means that,
besides the wind generation forecasts discussed in this paper,
additional predictions of potential generation losses (e.g., the
probability of failure of various equipment) are to be issued, for
instance, based on reliability models in the spirit of \citet
{Billinton84}. Forecasts for the electric load can in addition be
obtained from one of the numerous methods recently surveyed by \citet
{Hahn09}, though very few of them look at full predictive densities.

Assuming independence of the various random variables, the overall
uncertainty, represented by a predictive marginal density $\hat
{f}^{O}_{t+k|t}$, is obtained through convolution,
%
%
%
\begin{equation}
\hat{f}^{O}_{t+k|t} = \hat{f}^{\varepsilon_L}_{t+k|t}
\ast\hat{f}^{G}_{t+k|t} \ast\hat{f}^{\varepsilon_Y}_{t+k|t}.
\label{eq:ouncert}
\end{equation}
This predictive density is split into its positive and negative parts,
yielding $\hat{f}^{O^+}_{t+k|t}$ and $\hat{f}^{O^-}_{t+k|t}$, since
decisions about downward and upward reserve capacities are to be made
separately.

After such a description of system-wide uncertainties, the system
operator can plug this density into a cost-loss analysis
\citep{Matos10}, similar in essence to the market participation
problem presented in the above. Based on cost functions $g^\downarrow$
and $g^\uparrow$ for the downward and upward cases, the optimal
amounts of reserve capacities (in an expected utility maximization
sense) are the solution of stochastic optimization problems of the
form
%
%
%
\begin{equation}
q^{\uparrow*}_{t+k} = \argmin_{q^\uparrow_{t+k}} \mathbb{E} \bigl[
g^\uparrow\bigl(O^-_{t+k}, q^\uparrow_{t+k} \bigr)
\bigr] \label{eq:optresup}
\end{equation}
and
%
%
%
\begin{equation}
q^{\downarrow*}_{t+k} = \argmin_{q^\downarrow_{t+k}} \mathbb{E}
\bigl[
g^\downarrow\bigl(O^+_{t+k}, q^\downarrow_{t+k} \bigr)
\bigr], \label{eq:optredown}
\end{equation}
which may be solved analytically or numerically, depending upon the
complexity of the cost functions. Here the optimal reserve levels
relate to specific quantiles of the predictive densities for the system
margin $O_{t+k}$. However, it would be difficult to link the optimal
reserve levels to specific quantiles of the input predictive densities
of wind power generation.

In its more complex form the reserve quantification problem requires
accounting for dependencies in space and in time, similar to the
trading problems, with many more considerations relating to operational
constraints, for example, unit characteristics (capability to increase
or decrease power output over a predefined period of time---so-called
ramping characteristics, nonconvexities in costs, etc.), and,
potentially, risk aversion. The resulting stochastic optimization
problems take the general form of those described and analyzed in
Ortega-Vazquez and\break  Kirschen (\citeyear{Ortega09}).
They require space--time trajectories for all input variables.

\section{Modeling and Forecasting Wind Power in a Probabilistic Framework}
\label{sec:mod}

Decision-making problems relating to an optimal management of wind
power generation in power systems and electricity markets require
different types of forecasts as input. The lead forecast range
considered in the above is between 13 and 37 hours ahead, with an
hourly temporal resolution for the forecasts. In practice, various
forecast ranges, spatial and temporal resolutions, are of relevance
depending upon the decision problem. For instance, the shorter lead
times, say, between 10 minutes and 2 hours ahead, are also crucial for
a number of dispatch and control problems. Below are presented the
leading forms of forecasts for wind power generation, as well as
example approaches to generate them.

\subsection{Point Predictions}
\label{ssec:pfore}

The traditional deterministic view of a large number of power system
operators translates to preferring single-valued forecasts. These
so-called point predictions are seen as easier to appraise and handle
at the time of making decisions.

When describing at time $t$ the random variable $\mathbf{Y}_{\mathbf
{s}, t+\mathbf{k}}$ of a set of locations $\mathbf{s}$ and lead times
$\mathbf{k}$, point forecasts comprise a summary value for each and
every marginal distribution of $\mathbf{Y}_{\mathbf{s},t+\mathbf{k}}$
in time and in space. Typically, if one aims at minimizing a quadratic
criterion (i.e., in a Least Squares sense), a point forecast for
location $s$ and lead time $k$ corresponds to the conditional
expectation for $Y_{s,t+k}$ given the information set available up to
time $t$, the chosen model and estimated parameters. With respect to a
predictive density $\hat{f}_{s,t+k|t}$ for location $s$ and lead time
$k$, that point forecast therefore corresponds to
%
%
%
\begin{equation}
\hat{y}_{s,t+k|t} = \int_0^1 y
\hat{f}_{s,t+k|t} ( y ) \,dy. \label{eq:pfore}
\end{equation}
Integration is between 0 and 1 since one is dealing with power values
normalized by the nominal capacity of the wind farm or group of wind
farms of interest.

To issue point predictions at time $t$, the forecaster utilizes an
information set $\Omega_t$, a set containing measurements $\Omega^o_t$
(including observations of power and of relevant meteorological
variables, the notation ``$^o$'' meaning ``observation'') over the area
covered, as well as meteorological forecasts $\Omega^f_t$ (with
``$^f$''
for ``forecast'') for these relevant variables, $\Omega_t \subseteq
\Omega
^o_t \cup\Omega^f_t$. Based on this wealth of available information,
different types of models of the general form
%
%
%
\begin{equation}
Y_{s,t+k} = h(\Omega_t) + \varepsilon_{s,t+k}
\label{eq:pforem}
\end{equation}
were proposed, where $\varepsilon_{s,t+k}$ is a noise term with zero
mean and finite variance.

Indeed, when focusing on a single location (a wind farm), it may be
that point forecasts can be issued in an inexpensive way based on local
measurements only, and in a linear time-series framework. The first
proposal in the literature is that of \citet{Brown84}, using
Auto-Regressive Moving Average (ARMA) models for wind speed
observations and for lead times between a few hours and a few days.
When focusing on wind power directly for very short range (say, for
lead times less than 2 hours), even simpler Auto-Regressive models of
order $p$, that is,
%
%
%
\begin{equation}
Y_{s,t+k} = \theta_0 + \sum_{i \in\mathcal{L}}
\theta_i Y_{s,t-i+1} + \sigma\varepsilon_{s,t+k},
\label{eq:ar}
\end{equation}
are difficult to outperform, possibly after data transformation \citep
{Pinson12}. In the above, $\mathcal{L} \subset\mathbb{N}^+$ is a set
of lag values of dimension $p$, while $\varepsilon_{s,t+k}$ is a
standard Gaussian noise term, scaled by a standard deviation value
$\sigma$. In addition, $k=1$ if the AR model is for 1-step ahead
prediction only, or to be used in an iterative fashion for $k$-step
ahead prediction, while $k>1$ if one uses the AR model for direct
$k$-step ahead forecasting.

These models were generalized by a few authors by accounting for
off-site observations and/or by accounting for regime-switching
dynamics of the time-series. A regime-switching version of the
model\break
in~\eqref{eq:ar} assumes different dynamic behaviors in the various
regimes, as expressed by
%
%
%
\begin{equation}\qquad
Y_{s,t+k} = \theta_0^{(r_t)} + \sum
_{i \in\mathcal{L}} \theta_i^{(r_t)} Y_{s,t-i+1} +
\sigma^{(r_t)} \varepsilon_{t+k}, \label{eq:regar}
\end{equation}
where $r_t$ is a realization at time $t$ of a regime sequence defined
by discrete random variables, with $r_t \in\{ 1,2,\ldots,R \}$,
$\forall t$, and $R$ is the number of regimes. The number of lags and
the noise variance may differ from one regime to another. The regime
sequence can be defined based on an observable process, like wind
direction at time $t$ or a previous wind power measurement, yielding
models of the Threshold Auto-Regressive (TAR) family, which are common
in econometrics \citep{Tong11}. As an example for wind speed modeling
and forecasting, \citet{Reikard08} proposed to consider temperature as
driving the regime-switch\-ing behavior in wind dynamics. In contrast,
the class of Markov-Switching Auto-Regressive (MSAR) models, also
popular in econometrics since the work of \citet{Hamilton89}, assumes
that the regime sequence relies on an unobservable process. MSAR models
were shown to be able to mimic the observable switching in wind power
dynamics, especially offshore, that cannot be explained by available
meteorological measurements \citep{Pinson12c}.

Incorporating off-site information in regime-switch\-ing models of the
form of~\eqref{eq:regar} was proposed by \citet{Gneiting06} and
subsequently in a more general form by \citet{Hering10}, when focusing
on a data set for the Columbia Basin of eastern Washington and Oregon
in the US. The model in the regime-switching space--time (RST) approach
originally proposed by \citet{Gneiting06} can be formulated as
%
%
%
\begin{eqnarray}\label{eq:regaroffsite}
Y_{s,t+k} &=& \theta_0^{(r_t)} + \sum
_{i \in\mathcal{L}} \theta_i^{(r_t)} Y_{s,t-i+1}
\nonumber
\\[-8pt]
\\[-8pt]
\nonumber
&&{}+
\sum_{s_j \in\mathcal{S}} \sum_{i \in\mathcal
{L}_j}
\nu_{j,i}^{(r_t)} Y_{s_j,t-i+1} + \sigma^{(r_t)}
\varepsilon_{t+k},
\end{eqnarray}
that is, in the form of a TARX model (TAR with exogenous variables),
where a set of terms is added to the regime-switching model of~\eqref
{eq:regar}, for observations at off-site locations $s_j \in\mathcal
{S}_X$ and for a set of lagged values $i \in\mathcal{L}_j$ at this
location. Such models allow considering advection and diffusion of
upstream information, but require extensive expert knowledge for
optimizing the model structure.

Conditional parametric AR (CP--AR) models are another natural
generalization of regime-switching models,
%
%
%
\begin{eqnarray}\label{eq:cpar}
Y_{s,t+k} &=& \theta_0(\mathbf{x}_t) + \sum
_{i \in\mathcal{L}} \theta_i (\mathbf{x}_t)
Y_{s,t-i+1}
\nonumber
\\[-8pt]
\\[-8pt]
\nonumber
&&{} + \sigma(\mathbf{x}_t) \varepsilon_{s,t+k},
\end{eqnarray}
where instead of considering various regimes with their own dynamics,
the AR coefficients are replaced by smooth functions of a vector (of
low dimension, say, less than 3) of an exogenous variable $\mathbf{x}$,
for instance, wind direction only in \citet{Pinson12}. The noise
variance can be seen as a function of $\mathbf{x}$, or as a constant,
for simplicity. CP--AR models are relevant when switches between dynamic
behaviors are not that clear. Meanwhile, they also require fairly large
data sets for estimation, which are more and more available today.
Their use is motivated by empirical investigations at various wind
farms, where it was observed that specific meteorological variables
(e.g., wind direction, atmospheric stability) can substantially impact
power generation dynamics and predictability in a smooth manner.

%
\begin{figure*}

\includegraphics{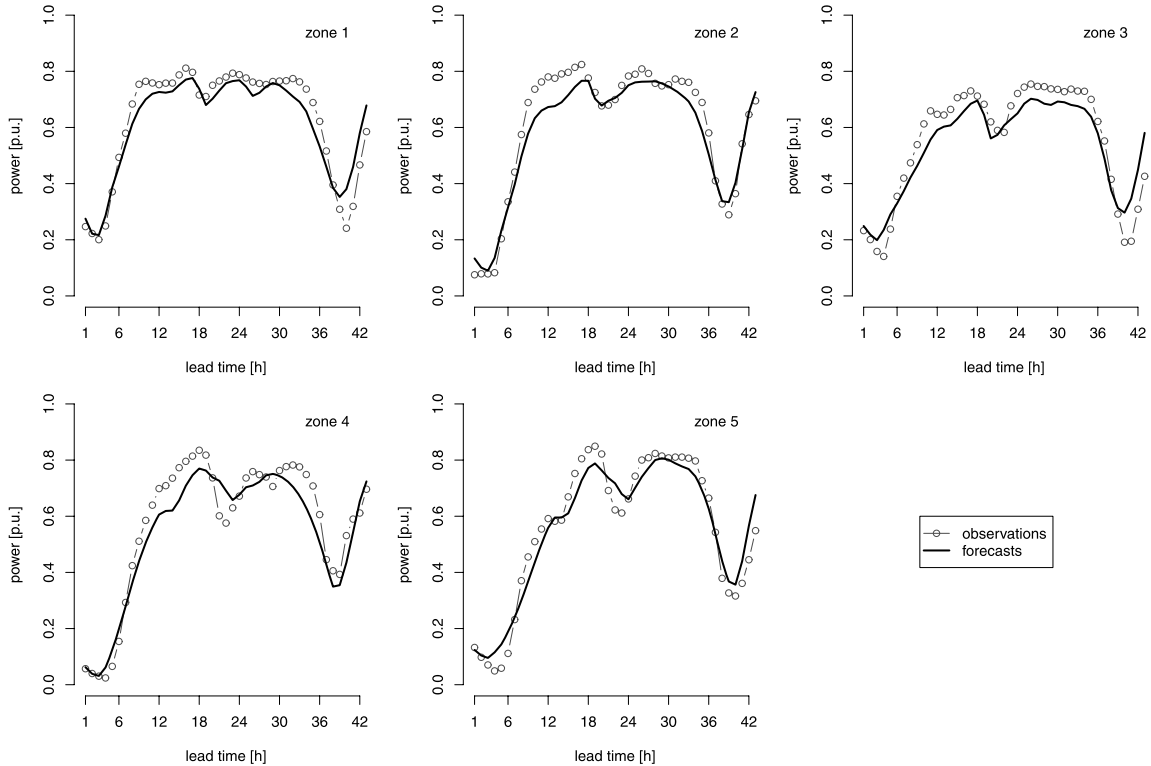}

\caption{Example episode with point forecasts for
the 5 aggregated zones of Western Denmark, as issued on 16 March 2007
at 06 UTC. Corresponding power measurements, obtained a posteriori, are
also shown.}\label{fig:pfs}
\end{figure*}

Other forms of conditional parametric models were proposed for further
lead times, also requiring additional meteorological forecasts as
input. As an example, a simplified version of the CP--ARX model (CP--AR
with exogenous variables) of \citet{Nielsen02} writes
%
%
%
\begin{eqnarray}
\label{eq:cparx} &&
Y_{s,t+k}\nonumber\\
&&\quad =  \theta^c_0(
\mathbf{x}_t) \cos\biggl(\frac{2\pi
h_{t+k}}{24} \biggr) +
\theta^s_0(\mathbf{x}_t) \sin\biggl(
\frac{2\pi
h_{t+k}}{24} \biggr)
\nonumber
\\[-8pt]
\\[-8pt]
\nonumber
&&\qquad{} + \theta_1(\mathbf{x}_t) Y_{s,t} +
\theta_2(\mathbf{x}_t) g(\hat{u}_{t+k|t},
\hat{v}_{t+k|t},k)\\
&&\qquad{} + \sigma\varepsilon_{s,t+k},
\nonumber
\end{eqnarray}
where $\hat{u}_{t+k|t}$ and $\hat{v}_{t+k|t}$ are forecasts of the wind
components (defining wind speed and direction) at the level of the wind
farm of interest. The vector $\mathbf{x}_t$ includes wind direction and
lead time. In addition, $g$ is used for a nonlinear conversion of the
information provided by meteorological forecasts to power generation,
for instance, modeled with nonparametric nonlinear regression (local
polynomial or spline-based). The model in~\eqref{eq:cparx} finally
includes diurnal Fourier harmonics for the correction of periodic
effects that may not be present in the meteorological forecasts, with
$h_{t+k}$ the hour of the day at lead time~$k$.

Besides~\eqref{eq:cparx}, a number of alternative approaches were
introduced in the past few years for predicting wind power generation
up to 2--3 days ahead based on both measurements and meteorological
forecasts. Notably, neural networks and other machine learning
approaches became popular after the original proposal of \citet
{Kariniotakis96} and more recently with the representative work of
\citet{Sideratos07}. For all of these models, parameters are commonly
estimated with Least Squares (LS) and Maximum Likelihood (ML)
approaches (and a Gaussian assumption for the residuals $\varepsilon
_{s,t+k}$), potentially made adaptive and recursive so as to allow for
smooth changes in the model parameters (accepting some form of
nonstationarity), while reducing computational costs. It was recently
argued that employing entropy-based criteria for parameter estimation
may be beneficial, as in \citet{Bessa09}, since they do not rely on any
assumption for the residual distributions. A more extensive review of
alternatives statistical approaches to point prediction of wind speed
and power can be found in \citet{Zhu12}.

As an illustration, Figure~\ref{fig:pfs} depicts example point
forecasts for the 5 aggregated zones of Western Denmark, issued on 16
March 2007 at 06 UTC based on the methodology described by \citet
{Nielsen02}. These have an hourly resolution up to 43 hours ahead, in
line with operational decision-making requirements. The well-captured
pattern for the first lead times originates from the combination of the
trend given by meteorological forecasts with the autoregressive
component based on local observational data. For the further lead
times, the dynamic wind power generation pattern is mainly driven by
the meteorological forecasts, though nonlinearly converted to power and
recalibrated to the specific conditions at these various aggregated zones.

In contrast with the introductory part of this section, where it was
mentioned that point forecasts corresponded to conditional expectation
estimates, \citeauthor{Gneiting10} (\citeyear{Gneiting10}, and references therein) discussed the
more general case of quantiles being optimal point forecasts in a
decision-theoretic framework. Indeed, in view of the operational
decision-making problems described in Section~\ref{sec:opchal}, it is
the case that if one accounts for the utility function of the
decision-makers at the time of issuing predictions, such forecasts
would then become specific quantiles,
%
%
%
\begin{equation}
\hat{y}_{s,t+k|t} = \hat{F}{}^{-1}_{s,t+k|t} (\alpha),
\label{eq:pforeq}
\end{equation}
whose nominal level $\alpha$ is determined from the utility function
and the structure of the problem itself. The information set and models
to be used for issuing quantile forecasts are similar in essence to
those for point predictions in the form of conditional expectations.
The estimation of model parameters is then performed based on the check
function criterion of \citet{Koenker78} or any general scoring rules
for quantiles \citep{Gneiting07}, instead of quadratic and
likelihood-based criteria. An example approach to point forecasting of
wind power generation where point forecasts actually are quantiles of
predictive densities is that of \citet{Moller08}, based on
time-adaptive quantile regression.

\subsection{Predictive Marginal Densities}

Point forecasts in the form of conditional expectations are somewhat
``just the mean of whatever may happen.'' These are not optimal inputs to
a large class of decision problems. Since the nominal level of quantile
forecasts to be used instead may vary in time while depending upon the
problem itself, or might be even unknown, issuing predictive densities
certainly is more relevant. Given the random variable $\mathbf
{Y}_{\mathbf{s},t+\mathbf{k}}$ whose characteristics are to be
predicted, these actually are predictive marginal densities $\hat{f}_{s,t+k|t}$ for all locations and lead times involved, individually,
with $\hat{F}_{s,t+k|t}$ the corresponding predictive c.d.f.s.

Today such a type of wind power forecast is issued in both parametric
and nonparametric frameworks. In the former case, based on an
assumption for the shape of predictive marginal densities (for
instance, motivated by an empirical investigation), one has
%
%
%
\begin{equation}
\hat{f}_{s,t+k|t} = f(y_{s,t+k};\hat{\bolds{\theta}}_{s,t+k|t}), \label{eq:parampfor}
\end{equation}
where $f$ is the density function for power to be generated at location
$s$ and time $t+k$, for the chosen probability distribution, for
example, truncated/\break censored Gaussian or Beta. In~\eqref{eq:parampfor},
$\hat{\bolds{\theta}}_{s,t+k|t}$ is the predicted value for the
vector of parameters fully characterizing that distribution, for
instance, a vector of parameters consisting of location and scale
parameters for the truncated/censored Gaussian and Beta distributions.
For these classes of distributions characterized by such limited sets
of parameters only, point forecasts as conditional expectations,
complemented by a variance estimator, for example, using exponential
smoothing or based on an ARCH/\break GARCH process, permit to directly obtain
location and scale parameters of predictive marginal densities. This
reliance on a limited number of parameters may be seen as desirable
since it eases subsequent estimation and related computational cost.

Models for the density parameters take a general form similar to that
in~\eqref{eq:pforem} (and subsequent models in Section~\ref{ssec:pfore}), that is, based on linear or nonlinear models with input
a subset $\Omega_t$ from the information set at time $t$. Example
parametric approaches include the RST approach of \citet{Gneiting06}
for predicting wind speed with truncated Gaussian distributions and
that of \citet{Pinson12} using Generalized Logit--Normal distributions
for wind power, also compared with censored Gaussian and Beta
assumptions. Similarly, \citet{Lau10} proposed employing Logit--Normal
distributions for aggregated wind power generation for the whole
Republic of Ireland.

In contrast, nonparametric approaches, since they do not rely on any
assumption for the shape of predictive densities, translate to focusing
on sets of quantile forecasts defining predictive c.d.f.s. These are
conveniently summarized by such sets of quantile forecasts,
%
%
%
\begin{eqnarray}\label{eq:densfor}
\quad&&\hat{F}_{s,t+k|t} = \bigl\{ \hat{q}_{s,t+k|t}^{(\alpha_i)} ; 0 \leq
\alpha_1 < \cdots< \alpha_i < \cdots
\nonumber
\\[-8pt]
\\[-8pt]
\nonumber
&&\hspace*{141pt}\quad<
\alpha_l \leq1 \bigr\},
\end{eqnarray}
with nominal levels $\alpha_i$ spread over the unit interval, though,
in practice, $\hat{F}_{s,t+k|t}$ is obtained by interpolation through
these sets of quantile forecasts. Actually, nonparametric approaches to
quantile forecasts may suffer from a limited number of relevant
observations for the very low and high nominal levels $\alpha$, say,
$\alpha,1-\alpha<0.05$, therefore potentially compromising the quality
of resulting forecasts. This was observed by \citet{Manganelli04} when
focusing on risk quantification approaches in finance, and more
particularly on dynamic quantile regression models for very low and
high levels. Even though the application of interest here is different,
the numerical aspects of estimating models for quantiles of wind power
generation for very low and high levels are similar. It may therefore
be advantageous under certain conditions to define nonparametric
predictive densities for their most central part, say, $\alpha
,1-\alpha
>0.05$, while parametric assumptions could be employed for the tails.

%
\begin{figure*}

\includegraphics{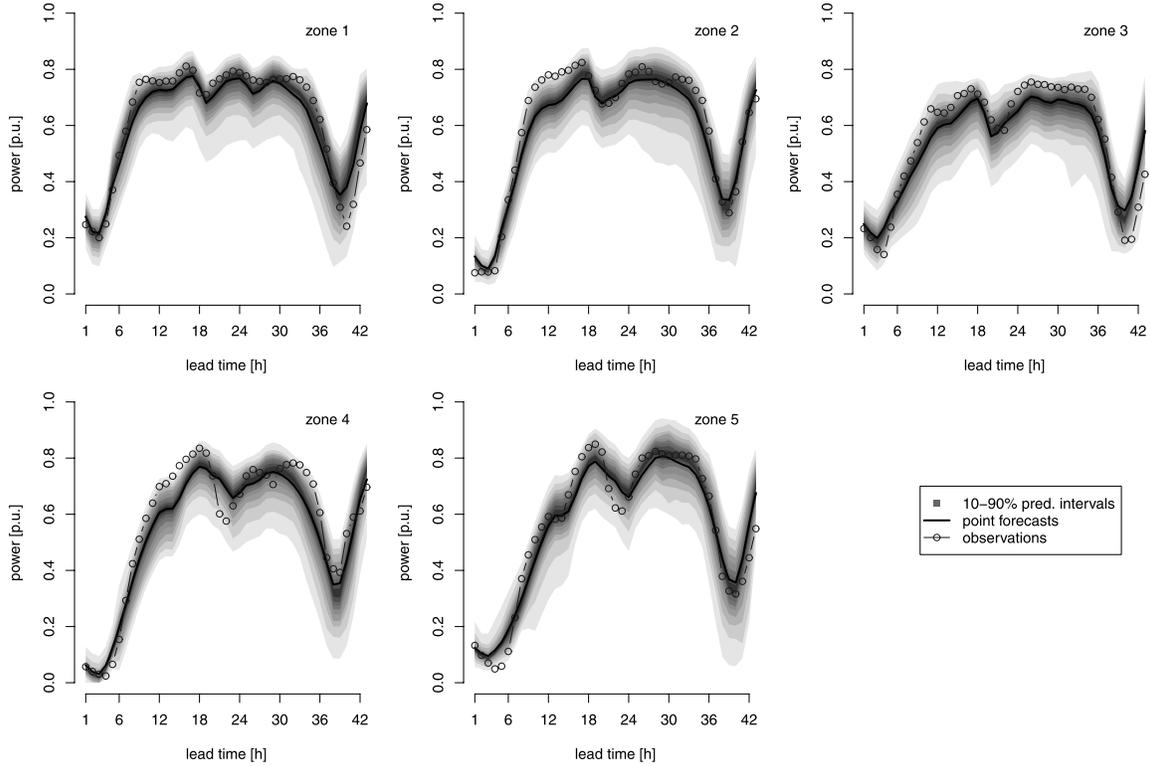}

\caption{Example episode with probabilistic
forecasts for the 5 aggregated zones of Western Denmark (and
corresponding measurements obtained a posteriori), as issued
on 16 March 2007 at 06 UTC. They take the form of so-called
river-of-blood fan charts [termed coined after Wallis (\citeyear{Wallis03})],
represented by a set of central prediction intervals with
increasing nominal coverage rates (from 10\% to 90\%).}\label
{fig:ppfs} 
\end{figure*}

A number of approaches for issuing nonparametric probabilistic
forecasts of wind power were proposed and benchmarked over the last
decade. In the most standard case, these are obtained from already
generated point predictions and, potentially, associated meteorological
forecasts. Maybe the most well-documented and widely applied methods
are the simple approach of \citet{Pinson10} consisting in dressing the
available point forecasts with predictive densities of forecast errors
made in similar conditions, as well as the local quantile regression of
\citet{Bremnes04} and the time-adaptive quantile regression of \citet
{Moller08}, to be used for each of the defining quantile forecasts. The
approach of \citet{Moller08} comprises an upgraded version of the
original proposal of \citet{Nielsen06}, where quantile forecasts of
wind power generation are conditional to previously issued point
forecasts and to input wind direction forecasts. As for point
predictions, neural network and machine learning techniques became
increasingly popular over the last few years for generating
nonparametric probabilistic predictions based on a set of
quantiles\break
\citep{Sideratos12}. In contrast to these methods using single-valued
forecasts of wind power and meteorological variables as input, a
relevant alternative relies on meteorological ensemble predictions,
that is, sets of multivariate space--time trajectories for
meteorological variables as issued by meteorological institutes [see
\citet{Leutbecher08} and the references therein], which are then
transformed to the wind power space. Ensemble forecasts attempt at
dynamically representing uncertainties in meteorological forecasts (as
well as spatial, temporal and inter-variable dependencies), by jointly
accounting for misestimation in the initial state of the Atmosphere and
for parameter uncertainty in the model dynamics. To obtain
probabilistic forecasts of wind power generation from such
meteorological ensembles, conventional approaches combine nonlinear
regression and kernel dressing of the ensemble trajectories, as in the
alternative proposals of \citet{Roulston03}; \citet{Taylor09} and
\citet
{Pinson09b}. In a similar vain, a general method for the conversion of
probabilistic forecasts of wind speed to power based on stochastic
power curves, thus accounting for additional uncertainties in the
wind-to-power conversion process in a Bayesian framework, was recently
described by \citet{Jeon12}.

Example nonparametric forecasts are shown in Figure~\ref{fig:ppfs} for
the same period as in Figure~\ref{fig:pfs}, as obtained by applying the
method of \citet{Pinson10} to the already issued point predictions and
their input meteorological forecast information. The characteristics of
these predictive densities smoothly evolve as a function of a number of
factors, for example, lead time, geographical location, time of the
year and level of power generation (since nonlinear and bounded power
curves shape forecast uncertainty). By construction, and through
adaptive estimation, these predictive densities are probabilistically
calibrated, meaning that observed levels for the defining quantile
forecasts correspond to the nominal ones. This is a crucial property of
probabilistic forecasts to be used as input to decision problems such
as those of Section~\ref{sec:opchal}, since a probabilistic bias in the
forecasts would yield suboptimality of resulting operational decisions.
Actually, in addition, probabilistic calibration is also a prerequisite
for application of the methods described in the following in order to
generate trajectories.

\subsection{Spatio-Temporal Trajectories}
\label{ssec:sttraj}

Both point forecasts and predictive densities are suboptimal inputs to
decision-making when spatial and temporal dependencies are involved. It
is then required to fully describe the density of the spatio-temporal
process $\mathbf{Y}_{\mathbf{s},t+\mathbf{k}}$. Following a proposal by
(\citet{Pinson09}) for wind power and, more \mbox{recently}, by \citeauthor{Moller13}\break (\citeyear{Moller13})
for multiple meteorological variables, the probabilistic forecast $\hat
{F}_{\mathbf{s},t+\mathbf{k}|t}$ can be fully characterized under a
Gaussian copula by the predictive marginal c.d.f.s $\hat{F}_{s,t+k|t}$,
$\forall s,k$, and by a space--time covariance matrix $\hat{\mathbf
{C}}_t$ linking all locations and lead times. In that case, using
notation similar to that of \citet{Moller13},
%
%
%
\begin{eqnarray} \label{eq:fullpcdfs}
&&\hat{F}_{\mathbf{s},t+\mathbf{k}|t} (\mathbf{y}_{\mathbf
{s},t+\mathbf
{k}} | \hat{\mathbf{C}}_t)
\nonumber
\\[-8pt]
\\[-8pt]
\nonumber
&&\quad=\Phi_{mn} \bigl( \bigl\{\Phi^{-1} \bigl(
\hat{F}_{s,t+k|t} (y) \bigr) \bigr\}_{s,k} | \hat{
\mathbf{C}}_t \bigr),
\end{eqnarray}
where $\mathbf{y}_{\mathbf{s},t+\mathbf{k}}$ was defined in~\eqref
{eq:normy} and $\Phi$ is the c.d.f. of a standard Gaussian variable, while
$\Phi_{mn}$ is that for a multivariate Gaussian of dimension $m \times
n$. Going beyond the Gaussian copula simplification, one could envisage
employing more refined copulas, though at the expense of additional
complexity. The interested reader may find an extensive introduction to
copulas in \citet{Nelsen06}.

This type of construction of multivariate probabilistic forecasts for
wind power generation in space and in time has clear advantages.
Indeed, given that all predictive densities forming the marginal
densities are calibrated, it may be assumed that one deals with a
latent space--time Gaussian process consisting of successive
multivariate random variables $\mathbf{Z}_t$ (each of dimension $m
\times n$) with realizations given by
%
%
%
\begin{eqnarray}\label{ref:latGauss}
&&\mathbf{z}_t = \bigl\{ \Phi^{-1} \bigl(
\hat{F}_{s,t+k|t} (y_{s,t+k}) \bigr); \nonumber
\\[-8pt]
\\[-8pt]
\nonumber
&&\hspace*{6pt}\qquad
s=s_1,s_2,
\ldots,s_m,
 k=1,2,\ldots,n \bigr\}.
\end{eqnarray}
Consequently, this latent Gaussian process can be used for identifying
and estimating a suitable parametric space--time structure or,
alternatively, if $m \times n$ is low and the sample size large, for
the tracking of the nonparametric (sample) covariance structure, for
instance, using exponential smoothing.

%
\begin{figure*}

\includegraphics{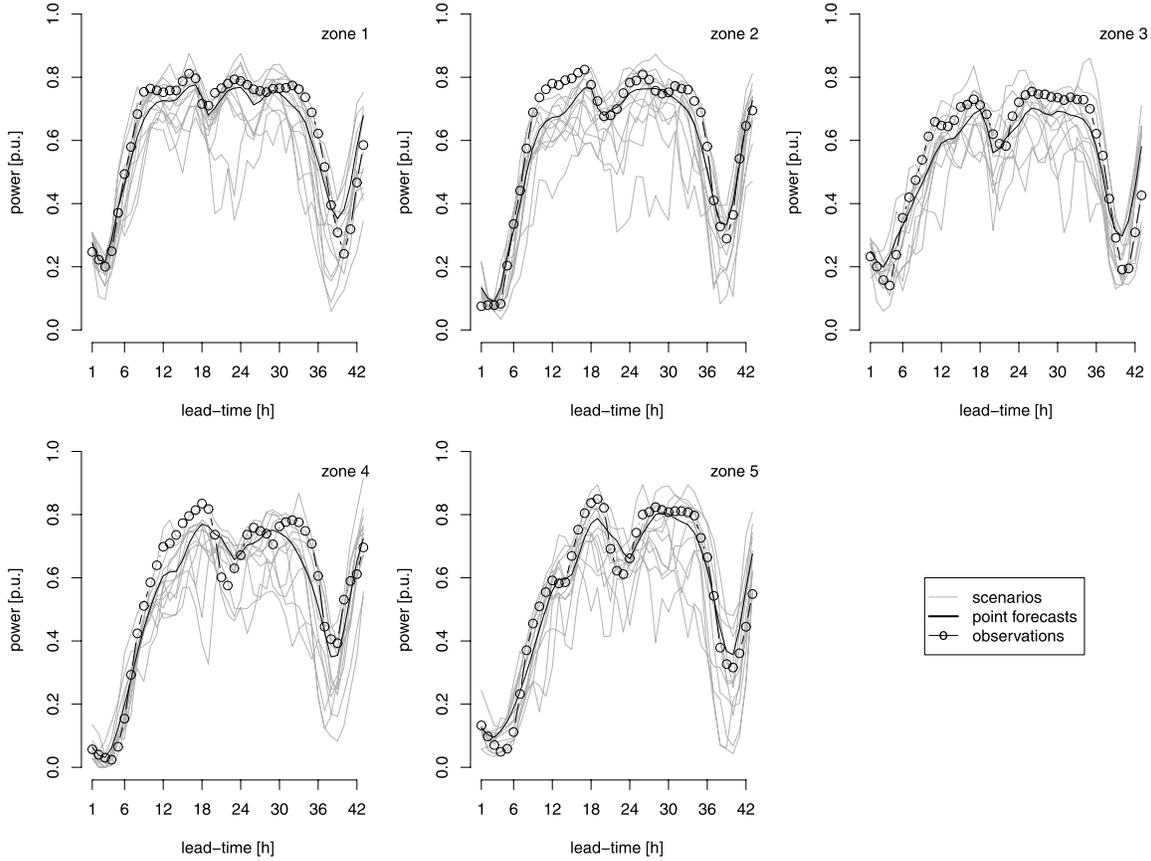}

\caption{Set of 12 space--time trajectories of wind power generation for
the 5 aggregated zones of Western Denmark, issued on the 16 March 2007
at 06 UTC.} \label{fig:pfsscen} 
\end{figure*}

Similarly, one of the advantages of this construction of multivariate
probabilistic forecasts based on a Gaussian copula is that it is fairly
straightforward to issue space--time trajectories. Remember that these
are the prime input to a large class of stochastic optimization
approaches, such as the advanced version of the problems presented in
Sections~\ref{ssec:bid} and~\ref{ssec:reserves}, where representation
of space--time interdependencies is required. Such trajectories also are
a convenient way to visualize the complex information conveyed by these
multivariate probabilistic forecasts, as hinted by \citet{Jorda10},
among others. Let us define by
%
%
%
\begin{eqnarray}
&&\hat{\mathbf{y}}_{\mathbf{s},t+\mathbf{k}|t}^{(j)} = \bigl\{\hat{y}{}^{(j)}_{s,t+k|t}; s=s_1,s_2,
\ldots,s_m,\nonumber\\
&&\hspace*{80pt}\qquad k=1,2,\ldots,n \bigr\}, \\
 \eqntext{j=1,2,\ldots,J ,}
\end{eqnarray}
a set of $J$ space--time trajectories issued at time $t$. As an
illustrative example, Figure~\ref{fig:pfsscen} gathers a set of $J=12$
space--time trajectories of wind power generation for the same episode
as in Figures~\ref{fig:pfs} and~\ref{fig:ppfs}. The covariance
structure $\hat{\mathbf{C}}_t$ used to fully specify the space--time
interdependence structure is obtained by exponential smoothing of the
sample covariance of the latent Gaussian process. The trajectories are
then obtained by first randomly sampling from a multivariate Gaussian
variable with the most up-to-date estimate of the space--time covariance
structure. Denote by $\mathbf{z}_t^{(j)}$ the $j$th sample,
whose components $z_{s,t+k}^{(j)}$ will directly relate to a location
$s$ and a lead time $k$ in the following. These multivariate Gaussian
samples are converted to wind power generation by a transformation
which is the inverse of that in~\eqref{ref:latGauss}. This yields
%
%
%
\begin{equation}
\qquad\hat{y}_{s,t+k}^{(j)} = \hat{F}{}^{-1}_{s,t+k|t}
\bigl( \Phi\bigl(z_{s,t+k}^{(j)} \bigr) \bigr)\quad \forall s,k,j .
\label{ref:invtransf}
\end{equation}
This type of visualization allows to appraise the temporal correlation
in wind power generation and potential forecast errors through
predictive densities, giving an extra level of information if compared
to the probabilistic forecasts of Figure~\ref{fig:ppfs}. There are
obvious limitations stemming from the dimensionality of the random
variable of interest. For instance, here, the spatial interdependence
structure, though serving to link these trajectories, is nearly
impossible to appreciate.


\section{Discussion: Upcoming Challenges}
\label{sec:concl}

Three decades of research in modeling and forecasting of power
generation from the wind have led to a solid understanding of the whole
chain from taking advantage of available meteorological and power
measurements, as well as meteorological forecasts, all the way to using
forecasts as input to decision-making. Today, methodologies are further
developed in a probabilistic framework, even though forecast users may
still prefer to be provided with single-valued predictions. Some
important challenges are currently under investigation or identified as
particularly relevant for the short to medium term. These are presented
and discussed below, with emphasis placed on new and better forecasts,
and forecast verification, as well as bridging the gap between forecast
quality and value.

\subsection{Improved Wind Power Forecasts: Extracting More out of the
Data}
\label{ssec:impfore}

Improving the quality of wind power forecasts is a constant challenge,
with strong expectations linked to the increased commitment of the
meteorological community to issue better forecasts of relevant weather
variables, mainly surface wind components. This will come, among other
things, from a better description of the physical phenomena involved,
especially in the boundary layer, as well as from an increased
resolution of the numerical schemes used to solve the systems of
partial differential equations.

Meanwhile, for statisticians, there are paths toward forecast
improvement that involve a better utilization of available measurement
data, combining measurements available on site and additional
observations spatially distributed around that site. Wind forecasts
used to issue power forecasts over a region seldom capture fully the
spatio-temporal dynamics of power generation owing to, for example, a
too coarse resolution (spatial and temporal) and timing errors with
respect to passages of weather fronts. However, all distributed
meteorological stations and wind turbines may serve as sensors in order
to palliate for these deficiencies. For the example of the Western
Denmark data set, \citet{Girard13} explored the spatio-temporal
characteristics of residuals after capturing local dynamics at all
individual sites, hinting at the role of prevailing weather conditions
on the space--time structure. For the same data set, \citet{Lau11}
investigated an anisotropic space--time covariance model based on a
Lagrangian approach, conditional to prevailing wind direction over the
region. Based on such analysis, it is required to propose nonlinear and
nonstationary spatio-temporal models for wind power generation, for
instance, using covariance structures conditional to prevailing weather
conditions, in the spirit of \citet{Huang04}. An advantage will be
that, instead of having to identify and estimate models for every
single site of interest (more than 400 for the Western Denmark data
set), and at various spatial and temporal resolutions of relevance to
forecast users, a single model would fit all purposes at once. Even
though more complex and potentially more costly in terms of parameter
estimation, they could lead to a substantial overall reduction in
computational time and expert knowledge necessary to set up and
maintain all individual models. Alternatively, approaches relying on
stochastic partial differential equations ought to be considered owing
to appealing features and recent advances in their linkage to
spatio-temporal covariance structures, as well as improved
computational solving \citep{Lindgren11}. Challenges\break  there, however,
relate to the complexity of the sto\-chastic processes involved,
requiring to account for the state-dependent diffusion part, and also
for\break  changes in the very dynamics of wind components, as induced by a
number of weather phenomena. It is not clear how all these aspects
could be accounted for in a compact set of stochastic differential
equations, which could be solved with existing numerical integration schemes.

The increasing availability of high-dimensional data sets, with a large
number of relevant meteorological and power systems variables, possibly
at high spatial and temporal resolutions, gives rise to a number of
challenges and opportunities related to data aggregation. These
challenges have already been identified in other fields, for example,
econometrics, where aggregation has shown its interest and potential
limitations. A relevant example work is that of \citet{Giacomini04},
which looks at the problem of pooling and forecasting spatially
correlated data sets. On the one hand, considering different levels of
aggregation for the wind power forecasting problem can permit to ease
the modeling task, by identifying groups of turbines with similar
dynamic behavior which could be modeled jointly. On the other hand,
this would lower the computational burden by reducing model size and
complexity. Proposals related to aggregation should, however, fully
consider the meteorological aspects at different temporal and spatial
scales, which may dynamically condition how aggregate models would be
representative of geographically distributed wind farms. One could
build on the classical Space--Time Auto-Regressive (STAR) model of
\citet
{Cliff75} by enhancing it to having dynamic and conditional space--time
covariance structures. In a similar vain, dynamic models for
spatio-temporal data such as those introduced by \citet{Stroud99} and
follow-up papers are appealing, since they provide an alternative
approach to data aggregation by seeing the overall spatial processes as
a linear combination of a limited number of local (polynomial) spatial
processes in the neighborhood of appropriately chosen locations.
Overall, various relevant directions to space--time modeling could be
explored, based on the substantial literature existing for other
processes and in other fields.

\subsection{New Forecast Methodologies and Forecast Products}
\label{ssec:newfore}

As a result of these efforts, new types of forecasts will be available
to decision-makers in the form of continuous surfaces and trajectories,
from which predictions with any spatial and temporal resolution could
be dynamically extracted. Similar to the development of meteorological
forecasting, the need for larger computational facilities might call
for centralizing efforts in generating and issuing wind power
predictions. Actually, in the opposite direction, a share of
practitioners request predictions of lower complexity that could be
better appraised by a\break  broader audience and more easily integrated into
existing operational processes. For instance, since accommodating the
variability of power fluctuations with successive periods of
fast-increasing and fast-decreasing power generation is seen as an
issue by some system operators in the US and in Australia,
methodologies were proposed for the prediction of so-called ramp
events, where the definition of these ``ramp events'' is based on the
very need of the decision-maker (\citeauthor{Bossavy13},\break  \citeyear{Bossavy13}; \citeauthor{Gallego13}, \citeyear{Gallego13}).

Besides, even though alternative parametric assumptions for predictive
marginal densities have\break  been analyzed and benchmarked, for example,
Beta (Bludszuweit, Dom{\'i}nguez-Navarro and
Llombart,\break  \citeyear{Bludszuweit08}), truncated and censored Gaussian, and
Generalized Logit--Normal \citep{Pinson12}, there is no clear
superiority of one over the others, for all potential lead times, level
of aggregation and wind dynamics themselves. This certainly originates
from the nonlinear and bounded curves representing the conversion of
wind to power, known to shape predictive densities. Such curves may
additionally be time-varying,\vadjust{\goodbreak} uncertain and conditional on various
external factors. This is why future work should consider these curves
as stochastic power curves, also described by multivariate
distributions, as a generalization of the proposal of \citet{Jeon12}.
Their impact on the shape of predictive densities ought to be better
understood. Then, combined with probabilistic forecasts of relevant
explanatory variables, for instance, from recalibrated meteorological
ensembles, stochastic power curves would naturally yield probabilistic
predictions of wind power generation, in a Bayesian framework. This is
since stochastic power curves comprise models of the joint density of
meteorological variables and of corresponding wind power generation.
Predictive densities of wind power generation would then be obtained by
applying Bayes rule, that is, by passing probabilistic forecasts of
meteorological variables through such stochastic power curve models.

To broaden up, and since operational decision-making problems are based
on interdependent variables (power generation from different renewable
energy sources, electric load and potentially market variables),
multivariate probabilistic forecasts for relevant pairing, or for all
of them together, should be issued in the future, with the weather as
the common driver. Similar to the proposal of \citet{Moller13} for
multivariate probabilistic forecasts of meteorological variables, one
could generalize the space--time trajectories of Section~\ref{ssec:sttraj}
to a multivariate setup. Alternatives\break  should be thought
of, allowing to directly obtain such spatio-temporal and multivariate
predictions, instead of having to go through predictive marginal
densities first.

\subsection{Verifying Probabilistic Forecasts of Ever-Increasing
Dimensionality}

Forecast verification is a subtle exercise already for the most simple
case of dealing with point forecasts, to be based on the joint
distribution of forecasts and observations \citep{Murphy87}. Focus is
today on verifying forecasts in a probabilistic framework, for
instance, following the paradigm of \citet{Gneiting07b} originally
introduced for the univariate case, based on calibration and sharpness
of predictive marginal densities. The nonlinear and double-bounded
nature of the wind power stochastic process (possibly also a
discrete-continuous mixture) renders the verification of probabilistic
forecasts more complex, especially for their calibration. It generally
calls for an extensive reliability assessment conditional\vadjust{\goodbreak} on variables
known to impact the shape of predictive densities: level of power, wind
direction, etc. In addition, the benchmarking and comparison of
forecasting methods ought to account for sample size and correlation
issues, since evaluation sets often are of limited size (though of
increasing length now that some wind farms have been operating for a
long time), while correlation in forecast errors and other criteria
(skill score values, probability integral transform) is necessarily
present for forecasts with lead times further than one step ahead.
Verifying high-dimensional forecasts, like space--time trajectories in
the most extreme case, based on small samples will necessarily yield
score values that may not fully reflect actual forecast quality even
though the score used is proper. Indeed, the deviations from the
expected score value, which could be estimated better with larger
samples, would be substantial. Correlation issues may only magnify this
problem, since they somewhat reduce the effective sample size for
estimation. An illustration of the combined effects of sampling and
correlation on the verification of probabilistic forecasts can be found
in \citet{Pinson10b}.

Going from univariate to multivariate aspects,\break  \citet{Gneiting08}
explained how the previously introduced paradigm can be readily
generalized for multivariate probabilistic predictions, yielding
an\break
evaluation framework including skill scores and diagnostic tools. An
application to the verification of temporal trajectories of wind power
generation in \citet{Pinson12b} illustrated its potential limitations
stemming from the high-dimensionality (there, $n=43$ lead times) of the
underlying random variables. Following the discussion in Section~\ref{ssec:impfore}, it is clear that new views on forecast verification
ought to be introduced and evaluated as dimensionality increases. For
instance, recent work by \citet{Hering11} showed how to compare spatial
predictions in a framework inspired by the Diebold--Mariano test and
with limited assumptions on the spatial processes themselves, thus
permitting to deal with high-dimensional predictions by focusing on
their spatial structure.

\subsection{Bridging the Gap Between Forecast Quality and Value}

\citet{Murphy93} introduced 3 types of goodness for weather forecasts,
also valid and relevant for other types of predictions like for wind
power. Out of these~3, quality and value play a particular role: {
(i)} quality relates\vadjust{\goodbreak} to the objective assessment of how well forecasts
describe the stochastic process of interest (and its realizations),
regardless of how the forecasts may be used subsequently, while {
(ii)} value corresponds to the economic/operational gain from
considering forecasts at the decision-making stage. Through the
introduction of representative operational decision problems in
Section~\ref{sec:opchal}, it was shown that optimal \mbox{forecasts} as input
to decision-making in a stochastic optimization framework take the form
of quantiles, predictive marginal densities or, finally, trajectories
describing the full spatio-temporal process. However, it is not clear
today how improving the quality of these forecasts, for instance, in
terms of reduced skill score values or increased probabilistic
calibration, may lead to added value for the decision-makers,
especially when they might use these forecasts sub-optimally. In
practice, this will call for more analytic work in a decision-theoretic
framework, by better linking skill scores of the forecasters and
utility of the decision-makers, as well as for a number of simulation
studies in order to simulate the usage of forecasts of varying quality
as input to a wide range of relevant operational problems. Full
benefits from integrating wind power generation into existing power
systems and through electricity markets will only be obtained by
optimally integrating forecasts in decision-making.

\section*{Acknowledgments}
The author was supported by the EU Commission through the project SafeWind
(ENK7-CT2008-213740) and the Danish
Public Service Obligation (PSO) fund under Radar\@Sea (contract no. 2009-1-0226), as well as the
Danish Strategic Research\break  \mbox{Council} under `5s'---Future Electricity Markets (12-132636/DSF), which are hereby acknowledged.
The author is also
grateful to ENFOR, DONG Energy, Vattenfall Denmark and
Energinet.dk for providing the data used in this paper. Acknowledgments
are finally due to Tilmann Gneiting
(Heidelberg University), Adrian Raftery (University of Washington) and
Patrick McSharry (University of Oxford)
for rows of inspiring discussions on probabilistic forecasting and
forecast verification, as well as Julija Tastu,
Pierre-Julien Trombe, Jan K. M{\o}ller and Henrik Madsen, among
others, at the Technical University of Denmark,
for contributing to broadening our understanding of spatio-temporal
dynamics and uncertainties in wind power
modeling and forecasting. Acknowledgments are finally due to two
reviewers and a guest editor for their comments and suggestions.
%

%



\begin{thebibliography}{84}

\bibitem[\protect\citeauthoryear{Ackermann}{2012}]{Ackermann12}
%
\begin{bbook}[auto:STB|2013/09/19|12:14:10]
\bauthor{\bsnm{Ackermann},~\bfnm{T.}\binits{T.}}
(\byear{2012}).
\btitle{Wind Power in Power Systems},
\bedition{2nd} ed.
\bpublisher{Wiley}, \blocation{Chichester}.
\bptok{imsref}%
\end{bbook}
%
\endbibitem

\bibitem[\protect\citeauthoryear{Bessa, Miranda and Gama}{2009}]{Bessa09}
%
\begin{barticle}[auto:STB|2013/09/19|12:14:10]
\bauthor{\bsnm{Bessa},~\bfnm{R.~J.}\binits{R.~J.}},
\bauthor{\bsnm{Miranda},~\bfnm{V.}\binits{V.}} \AND
\bauthor{\bsnm{Gama},~\bfnm{J.}\binits{J.}}
(\byear{2009}).
\btitle{Entropy and correntropy against minimum square error in
offline and
online three-day ahead wind power forecasting}.
\bjournal{IEEE T. Power Syst.}
\bvolume{24}
\bpages{1657--1666}.
\bptok{imsref}%
\end{barticle}
%
\endbibitem

\bibitem[\protect\citeauthoryear{Billinton and Allan}{1984}]{Billinton84}
%
\begin{bbook}[auto:STB|2013/09/19|12:14:10]
\bauthor{\bsnm{Billinton},~\bfnm{R.}\binits{R.}} \AND
\bauthor{\bsnm{Allan},~\bfnm{R.~N.}\binits{R.~N.}}
(\byear{1984}).
\btitle{Reliability Evaluation of Power Systems}.
\bpublisher{Plenum Press}, \blocation{London}.
\bptok{imsref}%
\end{bbook}
%
\endbibitem

\bibitem[\protect\citeauthoryear{Bludszuweit, Dom{\'i}nguez-Navarro and
Llombart}{2008}]{Bludszuweit08}
%
\begin{barticle}[auto:STB|2013/09/19|12:14:10]
\bauthor{\bsnm{Bludszuweit},~\bfnm{H.}\binits{H.}},
\bauthor{\bsnm{Dom{\'i}nguez-Navarro},~\bfnm{J.~A.}\binits{J.~A.}}
\AND
\bauthor{\bsnm{Llombart},~\bfnm{A.}\binits{A.}}
(\byear{2008}).
\btitle{Statistical analysis of wind power forecast errors}.
\bjournal{IEEE T. Power Syst.}
\bvolume{23}
\bpages{983--991}.
\bptok{imsref}%
\end{barticle}
%
\endbibitem

\bibitem[\protect\citeauthoryear{Bossavy, Girard and
Kariniotakis}{2013}]{Bossavy13}
%
\begin{barticle}[auto:STB|2013/09/19|12:14:10]
\bauthor{\bsnm{Bossavy},~\bfnm{A.}\binits{A.}},
\bauthor{\bsnm{Girard},~\bfnm{R.}\binits{R.}} \AND
\bauthor{\bsnm{Kariniotakis},~\bfnm{G.}\binits{G.}}
(\byear{2013}).
\btitle{Forecasting ramps of wind power production with numerical weather
prediction ensembles}.
\bjournal{Wind Energy}
\bvolume{16}
\bpages{51--63}.
\bptok{imsref}%
\end{barticle}
%
\endbibitem

\bibitem[\protect\citeauthoryear{Botterud et~al.}{2010}]{Botterud10}
%
\begin{barticle}[auto:STB|2013/09/19|12:14:10]
\bauthor{\bsnm{Botterud},~\bfnm{A.}\binits{A.}},
\bauthor{\bsnm{Wang},~\bfnm{J.}\binits{J.}},
\bauthor{\bsnm{Miranda},~\bfnm{V.}\binits{V.}} \AND
\bauthor{\bsnm{Bessa},~\bfnm{R.~J.}\binits{R.~J.}}
(\byear{2010}).
\btitle{Wind power forecasting in US electricity markets}.
\bjournal{The Electricity Journal}
\bvolume{23}
\bpages{71--82}.
\bptok{imsref}%
\end{barticle}
%
\endbibitem

\bibitem[\protect\citeauthoryear{Bremnes}{2004}]{Bremnes04}
%
\begin{barticle}[auto:STB|2013/09/19|12:14:10]
\bauthor{\bsnm{Bremnes},~\bfnm{J.~B.}\binits{J.~B.}}
(\byear{2004}).
\btitle{Probabilistic wind power forecasts using local quantile regression}.
\bjournal{Wind Energy}
\bvolume{7}
\bpages{47--54}.
\bptok{imsref}%
\end{barticle}
%
\endbibitem

\bibitem[\protect\citeauthoryear{Brown, Katz and Murphy}{1984}]{Brown84}
%
\begin{barticle}[auto:STB|2013/09/19|12:14:10]
\bauthor{\bsnm{Brown},~\bfnm{B.~G.}\binits{B.~G.}},
\bauthor{\bsnm{Katz},~\bfnm{R.~W.}\binits{R.~W.}} \AND
\bauthor{\bsnm{Murphy},~\bfnm{A.~M.}\binits{A.~M.}}
(\byear{1984}).
\btitle{Time series models to simulate and forecast wind speed and
wind power}.
\bjournal{J. Appl. Meteor.}
\bvolume{23}
\bpages{1184--1195}.
\bptok{imsref}%
\end{barticle}
%
\endbibitem

\bibitem[\protect\citeauthoryear{Cliff and Ord}{1984}]{Cliff75}
%
\begin{barticle}[auto:STB|2013/09/19|12:14:10]
\bauthor{\bsnm{Cliff},~\bfnm{A.~D.}\binits{A.~D.}} \AND
\bauthor{\bsnm{Ord},~\bfnm{J.~K.}\binits{J.~K.}}
(\byear{1984}).
\btitle{Space--time modeling with applications to regional forecasting}.
\bjournal{Trans. Inst. Br. Geogr.}
\bvolume{66}
\bpages{119--128}.
\bptok{imsref}%
\end{barticle}
%
\endbibitem

\bibitem[\protect\citeauthoryear{Conejo, Carrion and
Morales}{2010}]{Conejo10}
%
\begin{bmisc}[auto:STB|2013/09/19|12:14:10]
\bauthor{\bsnm{Conejo},~\bfnm{A.}\binits{A.}},
\bauthor{\bsnm{Carrion},~\bfnm{M.}\binits{M.}} \AND
\bauthor{\bsnm{Morales},~\bfnm{J.~M.}\binits{J.~M.}}
(\byear{2010}).
\bhowpublished{\textit{Decision-Making under Uncertainty in Electricity Markets}.
Springer, New York}.
\bptok{imsref}%
\end{bmisc}
%
\endbibitem

\bibitem[\protect\citeauthoryear{Conradsen, Nielsen and
Prahm}{1984}]{Conradsen84}
%
\begin{barticle}[auto:STB|2013/09/19|12:14:10]
\bauthor{\bsnm{Conradsen},~\bfnm{K.}\binits{K.}},
\bauthor{\bsnm{Nielsen},~\bfnm{L.~B.}\binits{L.~B.}} \AND
\bauthor{\bsnm{Prahm},~\bfnm{L.~P.}\binits{L.~P.}}
(\byear{1984}).
\btitle{Review of Weibull statistics for estimation of wind speed
distributions}.
\bjournal{J. Appl. Meteor.}
\bvolume{23}
\bpages{1173--1183}.
\bptok{imsref}%
\end{barticle}
%
\endbibitem

\bibitem[\protect\citeauthoryear{Doherty and O'Malley}{2005}]{Doherty05}
%
\begin{barticle}[auto:STB|2013/09/19|12:14:10]
\bauthor{\bsnm{Doherty},~\bfnm{R.}\binits{R.}} \AND
\bauthor{\bsnm{O'Malley},~\bfnm{M.}\binits{M.}}
(\byear{2005}).
\btitle{A new approach to quantify reserve demand in systems with significant
installed wind capacity}.
\bjournal{IEEE T. Power Syst.}
\bvolume{20}
\bpages{587--595}.
\bptok{imsref}%
\end{barticle}
%
\endbibitem

\bibitem[\protect\citeauthoryear{Gallego et~al.}{2013}]{Gallego13}
%
\begin{barticle}[auto:STB|2013/09/19|12:14:10]
\bauthor{\bsnm{Gallego},~\bfnm{C.}\binits{C.}},
\bauthor{\bsnm{Costa},~\bfnm{A.}\binits{A.}},
\bauthor{\bsnm{Cuerva},~\bfnm{A.}\binits{A.}},
\bauthor{\bsnm{Landberg},~\bfnm{L.}\binits{L.}},
\bauthor{\bsnm{Greaves},~\bfnm{B.}\binits{B.}} \AND
\bauthor{\bsnm{Collins},~\bfnm{J.}\binits{J.}}
(\byear{2013}).
\btitle{A wavelet-based approach for large wind power ramp characterisation}.
\bjournal{Wind Energy}
\bvolume{16}
\bpages{257--278}.
\bptok{imsref}%
\end{barticle}
%
\endbibitem

\bibitem[\protect\citeauthoryear{Giacomini and Granger}{2004}]{Giacomini04}
%
\begin{barticle}[mr]
\bauthor{\bsnm{Giacomini},~\bfnm{Raffaella}\binits{R.}} \AND
\bauthor{\bsnm{Granger},~\bfnm{Clive W.~J.}\binits{C.~W.~J.}}
(\byear{2004}).
\btitle{Aggregation of space--time processes}.
\bjournal{J. Econometrics}
\bvolume{118}
\bpages{7--26}.
\bid{doi={10.1016/S0304-4076(03)00132-5}, issn={0304-4076}, mr={2030964}}
\bptok{imsref}%
\end{barticle}
%
\endbibitem

\bibitem[\protect\citeauthoryear{Giebel et~al.}{2011}]{Giebel11}
%
\begin{bmisc}[auto:STB|2013/09/19|12:14:10]
\bauthor{\bsnm{Giebel},~\bfnm{G.}\binits{G.}},
\bauthor{\bsnm{Brownsword},~\bfnm{R.}\binits{R.}},
\bauthor{\bsnm{Kariniotakis},~\bfnm{G.}\binits{G.}},
\bauthor{\bsnm{Denhard},~\bfnm{M.}\binits{M.}} \AND
\bauthor{\bsnm{Draxl},~\bfnm{C.}\binits{C.}}
(\byear{2011}).
\bhowpublished{The state-of-the-art in short-term prediction of wind
power: A
literature overview, 2nd ed. Technical report. Available at
\href{http://www.orbit.dtu.dk}{orbit.dtu.dk}}.
\bptok{imsref}%
\end{bmisc}
%
\endbibitem

\bibitem[\protect\citeauthoryear{Girard and Allard}{2013}]{Girard13}
%
\begin{bmisc}[auto:STB|2013/09/19|12:14:10]
\bauthor{\bsnm{Girard},~\bfnm{R.}\binits{R.}} \AND
\bauthor{\bsnm{Allard},~\bfnm{D.}\binits{D.}}
(\byear{2013}).
\bhowpublished{Spatio-temporal propagation of wind power forecast errors.
\textit{Wind Energy}. To appear}.
\bptok{imsref}%
\end{bmisc}
%
\endbibitem

\bibitem[\protect\citeauthoryear{Gneiting}{2010}]{Gneiting10}
%
\begin{barticle}[auto:STB|2013/09/19|12:14:10]
\bauthor{\bsnm{Gneiting},~\bfnm{T.}\binits{T.}}
(\byear{2010}).
\btitle{Quantiles as optimal point forecasts}.
\bjournal{Int. J. Forecasting}
\bvolume{27}
\bpages{197--207}.
\bptok{imsref}%
\end{barticle}
%
\endbibitem

\bibitem[\protect\citeauthoryear{Gneiting, Balabdaoui and
Raftery}{2007}]{Gneiting07b}
%
\begin{barticle}[mr]
\bauthor{\bsnm{Gneiting},~\bfnm{Tilmann}\binits{T.}},
\bauthor{\bsnm{Balabdaoui},~\bfnm{Fadoua}\binits{F.}} \AND
\bauthor{\bsnm{Raftery},~\bfnm{Adrian~E.}\binits{A.~E.}}
(\byear{2007}).
\btitle{Probabilistic forecasts, calibration and sharpness}.
\bjournal{J. R. Stat. Soc. Ser. B Stat. Methodol.}
\bvolume{69}
\bpages{243--268}.
\bid{doi={10.1111/j.1467-9868.2007.00587.x}, issn={1369-7412}, mr={2325275}}
\bptok{imsref}%
\end{barticle}
%
\endbibitem

\bibitem[\protect\citeauthoryear{Gneiting and Raftery}{2007}]{Gneiting07}
%
\begin{barticle}[mr]
\bauthor{\bsnm{Gneiting},~\bfnm{Tilmann}\binits{T.}} \AND
\bauthor{\bsnm{Raftery},~\bfnm{Adrian~E.}\binits{A.~E.}}
(\byear{2007}).
\btitle{Strictly proper scoring rules, prediction, and estimation}.
\bjournal{J. Amer. Statist. Assoc.}
\bvolume{102}
\bpages{359--378}.
\bid{doi={10.1198/016214506000001437}, issn={0162-1459}, mr={2345548}}
\bptok{imsref}%
\end{barticle}
%
\endbibitem

\bibitem[\protect\citeauthoryear{Gneiting et~al.}{2006}]{Gneiting06}
%
\begin{barticle}[mr]
\bauthor{\bsnm{Gneiting},~\bfnm{Tilmann}\binits{T.}},
\bauthor{\bsnm{Larson},~\bfnm{Kristin}\binits{K.}},
\bauthor{\bsnm{Westrick},~\bfnm{Kenneth}\binits{K.}},
\bauthor{\bsnm{Genton},~\bfnm{Marc~G.}\binits{M.~G.}} \AND
\bauthor{\bsnm{Aldrich},~\bfnm{Eric}\binits{E.}}
(\byear{2006}).
\btitle{Calibrated probabilistic forecasting at the stateline wind energy
center:\vadjust{\goodbreak} The regime-switching space--time method}.
\bjournal{J. Amer. Statist. Assoc.}
\bvolume{101}
\bpages{968--979}.
\bid{doi={10.1198/016214506000000456}, issn={0162-1459}, mr={2324108}}
\bptok{imsref}%
\end{barticle}
%
\endbibitem

\bibitem[\protect\citeauthoryear{Gneiting et~al.}{2008}]{Gneiting08}
%
\begin{barticle}[mr]
\bauthor{\bsnm{Gneiting},~\bfnm{Tilmann}\binits{T.}},
\bauthor{\bsnm{Stanberry},~\bfnm{Larissa~I.}\binits{L.~I.}},
\bauthor{\bsnm{Grimit},~\bfnm{Eric~P.}\binits{E.~P.}},
\bauthor{\bsnm{Held},~\bfnm{Leonhard}\binits{L.}} \AND
\bauthor{\bsnm{Johnson},~\bfnm{Nicholas~A.}\binits{N.~A.}}
(\byear{2008}).
\btitle{Assessing probabilistic forecasts of multivariate quantities,
with an
application to ensemble predictions of surface winds}.
\bjournal{TEST}
\bvolume{17}
\bpages{211--235}.
\bid{doi={10.1007/s11749-008-0114-x}, issn={1133-0686}, mr={2434318}}
\bptok{imsref}%
\end{barticle}
%
\endbibitem

\bibitem[\protect\citeauthoryear{Hahn, Meyer-Nieberg and
Pickl}{2009}]{Hahn09}
%
\begin{barticle}[auto:STB|2013/09/19|12:14:10]
\bauthor{\bsnm{Hahn},~\bfnm{H.}\binits{H.}},
\bauthor{\bsnm{Meyer-Nieberg},~\bfnm{S.}\binits{S.}} \AND
\bauthor{\bsnm{Pickl},~\bfnm{S.}\binits{S.}}
(\byear{2009}).
\btitle{Electric load forecasting methods: Tools for decision making}.
\bjournal{European J. Oper. Res.}
\bvolume{199}
\bpages{902--907}.
\bptok{imsref}%
\end{barticle}
%
\endbibitem

\bibitem[\protect\citeauthoryear{Hamilton}{1989}]{Hamilton89}
%
\begin{barticle}[mr]
\bauthor{\bsnm{Hamilton},~\bfnm{James~D.}\binits{J.~D.}}
(\byear{1989}).
\btitle{A new approach to the economic analysis of nonstationary time series
and the business cycle}.
\bjournal{Econometrica}
\bvolume{57}
\bpages{357--384}.
\bid{doi={10.2307/1912559}, issn={0012-9682}, mr={0996941}}
\bptok{imsref}%
\end{barticle}
%
\endbibitem

\bibitem[\protect\citeauthoryear{Haslett and Raftery}{1989}]{Haslett89}
%
\begin{barticle}[auto:STB|2013/09/19|12:14:10]
\bauthor{\bsnm{Haslett},~\bfnm{J.}\binits{J.}} \AND
\bauthor{\bsnm{Raftery},~\bfnm{A.~E.}\binits{A.~E.}}
(\byear{1989}).
\btitle{Space--time modelling with long-memory dependence: Assessing Ireland's
wind power resource (with discussion)}.
\bjournal{J. R. Stat. Soc. Ser. C Appl. Stat.}
\bvolume{38}
\bpages{1--50}.
\bptok{imsref}%
\end{barticle}
%
\endbibitem

\bibitem[\protect\citeauthoryear{Hering and Genton}{2010}]{Hering10}
%
\begin{barticle}[mr]
\bauthor{\bsnm{Hering},~\bfnm{Amanda~S.}\binits{A.~S.}} \AND
\bauthor{\bsnm{Genton},~\bfnm{Marc~G.}\binits{M.~G.}}
(\byear{2010}).
\btitle{Powering up with space--time wind forecasting}.
\bjournal{J. Amer. Statist. Assoc.}
\bvolume{105}
\bpages{92--104}.
\bid{doi={10.1198/jasa.2009.ap08117}, issn={0162-1459}, mr={2757195}}
\bptok{imsref}%
\end{barticle}
%
\endbibitem

\bibitem[\protect\citeauthoryear{Hering and Genton}{2011}]{Hering11}
%
\begin{barticle}[mr]
\bauthor{\bsnm{Hering},~\bfnm{Amanda~S.}\binits{A.~S.}} \AND
\bauthor{\bsnm{Genton},~\bfnm{Marc~G.}\binits{M.~G.}}
(\byear{2011}).
\btitle{Comparing spatial predictions}.
\bjournal{Technometrics}
\bvolume{53}
\bpages{414--425}.
\bid{doi={10.1198/TECH.2011.10136}, issn={0040-1706}, mr={2850473}}
\bptok{imsref}%
\end{barticle}
%
\endbibitem

\bibitem[\protect\citeauthoryear{Huang and Hsu}{2004}]{Huang04}
%
\begin{barticle}[auto:STB|2013/09/19|12:14:10]
\bauthor{\bsnm{Huang},~\bfnm{H.~C.}\binits{H.~C.}} \AND
\bauthor{\bsnm{Hsu},~\bfnm{N.~J.}\binits{N.~J.}}
(\byear{2004}).
\btitle{Modeling transport effects on ground-level ozone using a non-stationary
space--time model}.
\bjournal{Environmetrics}
\bvolume{15}
\bpages{251--268}.
\bptok{imsref}%
\end{barticle}
%
\endbibitem

\bibitem[\protect\citeauthoryear{Jeon and Taylor}{2012}]{Jeon12}
%
\begin{barticle}[mr]
\bauthor{\bsnm{Jeon},~\bfnm{Jooyoung}\binits{J.}} \AND
\bauthor{\bsnm{Taylor},~\bfnm{James~W.}\binits{J.~W.}}
(\byear{2012}).
\btitle{Using conditional kernel density estimation for wind power density
forecasting}.
\bjournal{J.~Amer. Statist. Assoc.}
\bvolume{107}
\bpages{66--79}.
\bid{doi={10.1080/01621459.2011.643745}, issn={0162-1459}, mr={2949342}}
\bptok{imsref}%
\end{barticle}
%
\endbibitem

\bibitem[\protect\citeauthoryear{Jord{\`a} and Marcellino}{2010}]{Jorda10}
%
\begin{barticle}[mr]
\bauthor{\bsnm{Jord{\`a}},~\bfnm{{\`O}scar}\binits{{\`O}.}} \AND
\bauthor{\bsnm{Marcellino},~\bfnm{Massimiliano}\binits{M.}}
(\byear{2010}).
\btitle{Path forecast evaluation}.
\bjournal{J. Appl. Econometrics}
\bvolume{25}
\bpages{635--662}.
\bid{doi={10.1002/jae.1166}, issn={0883-7252}, mr={2758077}}
\bptok{imsref}%
\end{barticle}
%
\endbibitem

\bibitem[\protect\citeauthoryear{Joskow}{2011}]{Joskow11}
%
\begin{barticle}[auto:STB|2013/09/19|12:14:10]
\bauthor{\bsnm{Joskow},~\bfnm{P.~L.}\binits{P.~L.}}
(\byear{2011}).
\btitle{Comparing the costs of intermittent and dispatchable electricity
generating technologies}.
\bjournal{Amer. Econ. Rev.}
\bvolume{100}
\bpages{238--241}.
\bptok{imsref}%
\end{barticle}
%
\endbibitem

\bibitem[\protect\citeauthoryear{Kariniotakis, Stavrakakis and
Nogaret}{1996}]{Kariniotakis96}
%
\begin{barticle}[auto:STB|2013/09/19|12:14:10]
\bauthor{\bsnm{Kariniotakis},~\bfnm{G.~N.}\binits{G.~N.}},
\bauthor{\bsnm{Stavrakakis},~\bfnm{G.~S.}\binits{G.~S.}} \AND
\bauthor{\bsnm{Nogaret},~\bfnm{E.~F.}\binits{E.~F.}}
(\byear{1996}).
\btitle{Wind power forecasting using advanced neural networks models}.
\bjournal{IEEE T. Energy Conver.}
\bvolume{11}
\bpages{762--767}.
\bptok{imsref}%
\end{barticle}
%
\endbibitem

\bibitem[\protect\citeauthoryear{Koenker and Bassett}{1978}]{Koenker78}
%
\begin{barticle}[mr]
\bauthor{\bsnm{Koenker},~\bfnm{Roger}\binits{R.}} \AND
\bauthor{\bsnm{Bassett},~\bfnm{Gilbert}\binits{G.} \bsuffix{Jr.}}
(\byear{1978}).
\btitle{Regression quantiles}.
\bjournal{Econometrica}
\bvolume{46}
\bpages{33--50}.
\bid{issn={0012-9682}, mr={0474644}}
\bptok{imsref}%
\end{barticle}
%
\endbibitem

\bibitem[\protect\citeauthoryear{Landberg}{1999}]{Landberg99}
%
\begin{barticle}[auto:STB|2013/09/19|12:14:10]
\bauthor{\bsnm{Landberg},~\bfnm{L.}\binits{L.}}
(\byear{1999}).
\btitle{Short-term prediction of the power production from wind farms}.
\bjournal{J. Wind Eng. Ind. Aerodyn.}
\bvolume{80}
\bpages{207--220}.
\bptok{imsref}%
\end{barticle}
%
\endbibitem

\bibitem[\protect\citeauthoryear{Landberg and Watson}{1994}]{Landberg94}
%
\begin{barticle}[auto:STB|2013/09/19|12:14:10]
\bauthor{\bsnm{Landberg},~\bfnm{L.}\binits{L.}} \AND
\bauthor{\bsnm{Watson},~\bfnm{S.~J.}\binits{S.~J.}}
(\byear{1994}).
\btitle{Short-term prediction of local wind conditions}.
\bjournal{Bound.-Layer Meteor.}
\bvolume{70}
\bpages{171--195}.
\bptok{imsref}%
\end{barticle}
%
\endbibitem

\bibitem[\protect\citeauthoryear{Lange}{2005}]{Lange05}
%
\begin{barticle}[auto:STB|2013/09/19|12:14:10]
\bauthor{\bsnm{Lange},~\bfnm{M.}\binits{M.}}
(\byear{2005}).
\btitle{On the uncertainty of wind power predictions---Analysis of the forecast
accuracy and statistical distribution of errors}.
\bjournal{J. Solar Energ.-T. ASME}
\bvolume{127}
\bpages{177--184}.
\bptok{imsref}%
\end{barticle}
%
\endbibitem

\bibitem[\protect\citeauthoryear{Lange and Focken}{2006}]{Lange06}
%
\begin{bbook}[auto:STB|2013/09/19|12:14:10]
\bauthor{\bsnm{Lange},~\bfnm{M.}\binits{M.}} \AND
\bauthor{\bsnm{Focken},~\bfnm{U.}\binits{U.}}
(\byear{2006}).
\btitle{Physical Approach to Short-Term Wind Power Prediction}.
\bpublisher{Springer}, \blocation{Berlin}.
\bptok{imsref}%
\end{bbook}
%
\endbibitem

\bibitem[\protect\citeauthoryear{Lau}{2011}]{Lau11}
%
\begin{bmisc}[auto:STB|2013/09/19|12:14:10]
\bauthor{\bsnm{Lau},~\bfnm{A.}\binits{A.}}
(\byear{2011}).
\bhowpublished{Probabilistic wind power forecasts: From aggregated
approach to
spatiotemporal models. Ph.D. thesis, Univ. Oxford, Oxford, UK}.
\bptok{imsref}%
\end{bmisc}
%
\endbibitem

\bibitem[\protect\citeauthoryear{Lau and McSharry}{2010}]{Lau10}
%
\begin{barticle}[mr]
\bauthor{\bsnm{Lau},~\bfnm{Ada}\binits{A.}} \AND
\bauthor{\bsnm{McSharry},~\bfnm{Patrick}\binits{P.}}
(\byear{2010}).
\btitle{Approaches for multi-step density forecasts with application to
aggregated wind power}.
\bjournal{Ann. Appl. Stat.}
\bvolume{4}
\bpages{1311--1341}.
\bid{doi={10.1214/09-AOAS320}, issn={1932-6157}, mr={2758330}}
\bptok{imsref}%
\end{barticle}
%
\endbibitem

\bibitem[\protect\citeauthoryear{Letcher}{2008}]{Letcher08}
%
\begin{bbook}[auto:STB|2013/09/19|12:14:10]
\bauthor{\bsnm{Letcher},~\bfnm{T.~M.}\binits{T.~M.}}
(\byear{2008}).
\btitle{Future Energy: Improved, Sustainable and Clean Options for Our Planet}.
\bpublisher{Elsevier}, \blocation{Amsterdam}.
\bptok{imsref}%
\end{bbook}
%
\endbibitem

\bibitem[\protect\citeauthoryear{Leutbecher and Palmer}{2008}]{Leutbecher08}
%
\begin{barticle}[mr]
\bauthor{\bsnm{Leutbecher},~\bfnm{M.}\binits{M.}} \AND
\bauthor{\bsnm{Palmer},~\bfnm{T.~N.}\binits{T.~N.}}
(\byear{2008}).
\btitle{Ensemble forecasting}.
\bjournal{J. Comput. Phys.}
\bvolume{227}
\bpages{3515--3539}.
\bid{doi={10.1016/j.jcp.2007.02.014}, issn={0021-9991}, mr={2400226}}
\bptok{imsref}%
\end{barticle}
%
\endbibitem

\bibitem[\protect\citeauthoryear{Lindgren, Rue and
Lindstr{\"o}m}{2011}]{Lindgren11}
%
\begin{barticle}[mr]
\bauthor{\bsnm{Lindgren},~\bfnm{Finn}\binits{F.}},
\bauthor{\bsnm{Rue},~\bfnm{H{\aa}vard}\binits{H.}} \AND
\bauthor{\bsnm{Lindstr{\"o}m},~\bfnm{Johan}\binits{J.}}
(\byear{2011}).
\btitle{An explicit link between {G}aussian fields and {G}aussian {M}arkov
random fields: The stochastic partial differential equation approach}.
\bjournal{J.~R. Stat. Soc. Ser. B Stat. Methodol.}
\bvolume{73}
\bpages{423--498}.
\bid{doi={10.1111/j.1467-9868.2011.00777.x}, issn={1369-7412}, mr={2853727}}
\bptnote{check related}%
\bptok{imsref}%
\end{barticle}
%
\endbibitem

\bibitem[\protect\citeauthoryear{Manganelli and Engle}{2004}]{Manganelli04}
%
\begin{bincollection}[auto:STB|2013/09/19|12:14:10]
\bauthor{\bsnm{Manganelli},~\bfnm{S.}\binits{S.}} \AND
\bauthor{\bsnm{Engle},~\bfnm{R.~F.}\binits{R.~F.}}
(\byear{2004}).
\btitle{A comparison of value-at-risk models in finance}.
In \bbooktitle{Risk Measures for the 21st Century}
(\beditor{\bfnm{G.}\binits{G.}~\bsnm{Szeg{\"o}}}, ed.)
\bpages{123--143}.
\bpublisher{{W}iley}, \blocation{Chichester}.
\bptok{imsref}%
\end{bincollection}
%
\endbibitem

\bibitem[\protect\citeauthoryear{Matos and Bessa}{2010}]{Matos10}
%
\begin{barticle}[auto:STB|2013/09/19|12:14:10]
\bauthor{\bsnm{Matos},~\bfnm{M.~A.}\binits{M.~A.}} \AND
\bauthor{\bsnm{Bessa},~\bfnm{R.~J.}\binits{R.~J.}}
(\byear{2010}).
\btitle{Setting the operating reserve using probabilistic wind power
forecasts}.
\bjournal{IEEE T. Power Syst.}
\bvolume{26}
\bpages{594--603}.
\bptok{imsref}%
\end{barticle}
%
\endbibitem

\bibitem[\protect\citeauthoryear{M{\"o}ller, Lenkoski and
Thorarinsdottir}{2013}]{Moller13}
%
\begin{barticle}[auto:STB|2013/09/19|12:14:10]
\bauthor{\bsnm{M{\"o}ller},~\bfnm{A.}\binits{A.}},
\bauthor{\bsnm{Lenkoski},~\bfnm{A.}\binits{A.}} \AND
\bauthor{\bsnm{Thorarinsdottir},~\bfnm{T.~L.}\binits{T.~L.}}
(\byear{2013}).
\btitle{Multivariate probabilistic forecasting using ensemble Bayesian
model averaging and copulas}.
\bjournal{Q. J. Royal Met. Soc.}
\bvolume{139}
\bpages{982--991}.
\bptok{imsref}%
\end{barticle}
%
\endbibitem

\bibitem[\protect\citeauthoryear{M{\o}ller, Nielsen and Madsen}{2008}]{Moller08}
%
\begin{barticle}[mr]
\bauthor{\bsnm{M{\o}ller},~\bfnm{Jan~Kloppenborg}\binits{J.~K.}},
\bauthor{\bsnm{Nielsen},~\bfnm{Henrik~Aalborg}\binits{H.~A.}} \AND
\bauthor{\bsnm{Madsen},~\bfnm{Henrik}\binits{H.}}
(\byear{2008}).
\btitle{Time-adaptive quantile regression}.
\bjournal{Comput. Statist. Data Anal.}
\bvolume{52}
\bpages{1292--1303}.
\bid{doi={10.1016/j.csda.2007.06.027}, issn={0167-9473}, mr={2418566}}
\bptok{imsref}%
\end{barticle}
%
\endbibitem

\bibitem[\protect\citeauthoryear{Morthorst}{2003}]{Morthorst03}
%
\begin{barticle}[auto:STB|2013/09/19|12:14:10]
\bauthor{\bsnm{Morthorst},~\bfnm{P.~E.}\binits{P.~E.}}
(\byear{2003}).
\btitle{Wind power and the conditions at a liberalized power market}.
\bjournal{Wind Energy}
\bvolume{6}
\bpages{297--308}.
\bptok{imsref}%
\end{barticle}
%
\endbibitem

\bibitem[\protect\citeauthoryear{Mur~Amada and
Bayod~R{\'u}jula}{2010}]{MurAmada10}
%
\begin{bincollection}[auto:STB|2013/09/19|12:14:10]
\bauthor{\bsnm{Mur~Amada},~\bfnm{J.}\binits{J.}} \AND
\bauthor{\bsnm{Bayod~R{\'u}jula},~\bfnm{A.}\binits{A.}}
(\byear{2010}).
\btitle{Variability of wind and wind power}.
In \bbooktitle{Wind Power}
(\beditor{\bfnm{S.~M.}\binits{S.~M.}~\bsnm{Muyeen~Vukovar}}, ed.)
\bpages{289--320}.
\bpublisher{Intech Open}, \blocation{Rijeka, Croatia}.
\bptok{imsref}%
\end{bincollection}
%
\endbibitem

\bibitem[\protect\citeauthoryear{Murphy}{1993}]{Murphy93}
%
\begin{barticle}[auto:STB|2013/09/19|12:14:10]
\bauthor{\bsnm{Murphy},~\bfnm{A.~H.}\binits{A.~H.}}
(\byear{1993}).
\btitle{What is a good forecast? An essay on the nature of goodness in weather
forecasting}.
\bjournal{Wea. Forecasting}
\bvolume{8}
\bpages{281--293}.
\bptok{imsref}%
\end{barticle}
%
\endbibitem

\bibitem[\protect\citeauthoryear{Murphy and Winkler}{1987}]{Murphy87}
%
\begin{barticle}[auto:STB|2013/09/19|12:14:10]
\bauthor{\bsnm{Murphy},~\bfnm{A.~H.}\binits{A.~H.}} \AND
\bauthor{\bsnm{Winkler},~\bfnm{R.~L.}\binits{R.~L.}}
(\byear{1987}).
\btitle{A general framework for forecast verification}.
\bjournal{Mon. Wea. Rev.}
\bvolume{115}
\bpages{1330--1338}.
\bptok{imsref}%
\end{barticle}
%
\endbibitem

\bibitem[\protect\citeauthoryear{Nelsen}{2006}]{Nelsen06}
%
\begin{bbook}[mr]
\bauthor{\bsnm{Nelsen},~\bfnm{Roger~B.}\binits{R.~B.}}
(\byear{2006}).
\btitle{An Introduction to Copulas},
\bedition{2nd} ed.
\bpublisher{Springer}, \blocation{New York}.
\bid{mr={2197664}}
\bptok{imsref}%
\end{bbook}
%
\endbibitem

\bibitem[\protect\citeauthoryear{Nielsen}{2002}]{Nielsen02}
%
\begin{bmisc}[auto:STB|2013/09/19|12:14:10]
\bauthor{\bsnm{Nielsen},~\bfnm{T.~S.}\binits{T.~S.}}
(\byear{2002}).
\bhowpublished{Online prediction and control in nonlinear stochastic systems.
Ph.D. thesis, Dept. Informatics and Mathematical Modelling, Technical
Univ. Denmark}.
\bptok{imsref}%
\end{bmisc}
%
\endbibitem

\bibitem[\protect\citeauthoryear{Nielsen, Madsen and
Nielsen}{2006}]{Nielsen06}
%
\begin{barticle}[auto:STB|2013/09/19|12:14:10]
\bauthor{\bsnm{Nielsen},~\bfnm{H.~Aa.}\binits{H.~A.}},
\bauthor{\bsnm{Madsen},~\bfnm{H.}\binits{H.}} \AND
\bauthor{\bsnm{Nielsen},~\bfnm{T.~S.}\binits{T.~S.}}
(\byear{2006}).
\btitle{Using quantile regression to extend an existing wind power forecasting
system with probabilistic forecasts}.
\bjournal{Wind Energy}
\bvolume{9}
\bpages{95--108}.
\bptok{imsref}%
\end{barticle}
%
\endbibitem

\bibitem[\protect\citeauthoryear{Ortega-Vazquez and
Kirschen}{2009}]{Ortega09}
%
\begin{barticle}[auto:STB|2013/09/19|12:14:10]
\bauthor{\bsnm{Ortega-Vazquez},~\bfnm{M.~A.}\binits{M.~A.}} \AND
\bauthor{\bsnm{Kirschen},~\bfnm{D.~S.}\binits{D.~S.}}
(\byear{2009}).
\btitle{Estimating the spinning reserve requirements in systems with
significant wind power generation penetration}.
\bjournal{IEEE T. Power Syst.}
\bvolume{24}
\bpages{114--124}.
\bptok{imsref}%
\end{barticle}
%
\endbibitem

\bibitem[\protect\citeauthoryear{Palmer}{2012}]{Palmer12}
%
\begin{barticle}[auto:STB|2013/09/19|12:14:10]
\bauthor{\bsnm{Palmer},~\bfnm{T.~N.}\binits{T.~N.}}
(\byear{2012}).
\btitle{Towards the probabilistic Earth-system simulator: A vision for the
future of climate and weather prediction}.
\bjournal{Quart. J. Royal Met. Soc.}
\bvolume{665}
\bpages{841--\break861}.
\bptok{imsref}%
\end{barticle}
%
\endbibitem

\bibitem[\protect\citeauthoryear{Papavasiliou and Oren}{2013}]{Papavasiliou13}
\begin{barticle}[mr]
\bauthor{\bsnm{Papavasiliou},~\bfnm{Anthony}\binits{A.}} \AND
  \bauthor{\bsnm{Oren},~\bfnm{Shmuel~S.}\binits{S.~S.}}
(\byear{2013}).
\btitle{Multiarea stochastic unit commitment for high wind penetration in a
  transmission constrained network}.
\bjournal{Oper. Res.}
\bvolume{61}
\bpages{578--592}.
\bid{doi={10.1287/opre.2013.1174}, issn={0030-364X}, mr={3079732}}
\bptok{imsref}%
\end{barticle}
\endbibitem

\bibitem[\protect\citeauthoryear{Pe{\~n}a~Diaz, Gryning and
Mann}{2010}]{Pena10}
%
\begin{barticle}[auto:STB|2013/09/19|12:14:10]
\bauthor{\bsnm{Pe{\~n}a~Diaz},~\bfnm{A.}\binits{A.}},
\bauthor{\bsnm{Gryning},~\bfnm{S.~E.}\binits{S.~E.}} \AND
\bauthor{\bsnm{Mann},~\bfnm{J.}\binits{J.}}
(\byear{2010}).
\btitle{On the length-scale of the wind profile}.
\bjournal{Quart. J. Royal Met. Soc.}
\bvolume{136}
\bpages{2119--2131}.
\bptok{imsref}%
\end{barticle}
%
\endbibitem

\bibitem[\protect\citeauthoryear{Pinson}{2012}]{Pinson12}
%
\begin{barticle}[mr]
\bauthor{\bsnm{Pinson},~\bfnm{P.}\binits{P.}}
(\byear{2012}).
\btitle{Very-short-term probabilistic forecasting of wind power with
generalized logit--normal distributions}.
\bjournal{J. R. Stat. Soc. Ser. C Appl. Stat.}
\bvolume{61}
\bpages{555--576}.
\bid{doi={10.1111/j.1467-9876.2011.01026.x}, issn={0035-9254}, mr={2960738}}
\bptok{imsref}%
\end{barticle}
%
\endbibitem

\bibitem[\protect\citeauthoryear{Pinson, Chevallier and
Kariniotakis}{2007}]{Pinson07}
%
\begin{barticle}[auto:STB|2013/09/19|12:14:10]
\bauthor{\bsnm{Pinson},~\bfnm{P.}\binits{P.}},
\bauthor{\bsnm{Chevallier},~\bfnm{C.}\binits{C.}} \AND
\bauthor{\bsnm{Kariniotakis},~\bfnm{G.}\binits{G.}}
(\byear{2007}).
\btitle{Trading wind generation from short-term probabilistic
forecasts of wind
power}.
\bjournal{IEEE T. Power Syst.}
\bvolume{22}
\bpages{1148--1156}.
\bptok{imsref}%
\end{barticle}
%
\endbibitem

\bibitem[\protect\citeauthoryear{Pinson and Girard}{2012}]{Pinson12b}
%
\begin{barticle}[auto:STB|2013/09/19|12:14:10]
\bauthor{\bsnm{Pinson},~\bfnm{P.}\binits{P.}} \AND
\bauthor{\bsnm{Girard},~\bfnm{R.}\binits{R.}}
(\byear{2012}).
\btitle{Evaluating the quality of scenarios of short-term wind power
generation}.
\bjournal{Appl. Energ.}
\bvolume{96}
\bpages{12--20}.
\bptok{imsref}%
\end{barticle}
%
\endbibitem

\bibitem[\protect\citeauthoryear{Pinson and Kariniotakis}{2010}]{Pinson10}
%
\begin{barticle}[auto:STB|2013/09/19|12:14:10]
\bauthor{\bsnm{Pinson},~\bfnm{P.}\binits{P.}} \AND
\bauthor{\bsnm{Kariniotakis},~\bfnm{G.}\binits{G.}}
(\byear{2010}).
\btitle{Conditional prediction intervals of wind power generation}.
\bjournal{IEEE T. Power Syst.}
\bvolume{25}
\bpages{1845--1856}.
\bptok{imsref}%
\end{barticle}
%
\endbibitem

\bibitem[\protect\citeauthoryear{Pinson and Madsen}{2009}]{Pinson09b}
%
\begin{barticle}[auto:STB|2013/09/19|12:14:10]
\bauthor{\bsnm{Pinson},~\bfnm{P.}\binits{P.}} \AND
\bauthor{\bsnm{Madsen},~\bfnm{H.}\binits{H.}}
(\byear{2009}).
\btitle{Ensemble-based probabilistic forecasting at Horns Rev}.
\bjournal{Wind Energy}
\bvolume{12}
\bpages{137--155}.
\bptok{imsref}%
\end{barticle}
%
\endbibitem

\bibitem[\protect\citeauthoryear{Pinson and Madsen}{2012}]{Pinson12c}
%
\begin{barticle}[mr]
\bauthor{\bsnm{Pinson},~\bfnm{Pierre}\binits{P.}} \AND
\bauthor{\bsnm{Madsen},~\bfnm{Henrik}\binits{H.}}
(\byear{2012}).
\btitle{Adaptive modelling and forecasting of offshore wind power fluctuations
with {M}arkov-switching autoregressive models}.
\bjournal{J. Forecast.}
\bvolume{31}
\bpages{281--313}.
\bid{doi={10.1002/for.1194}, issn={0277-6693}, mr={2924797}}
\bptok{imsref}%
\end{barticle}
%
\endbibitem

\bibitem[\protect\citeauthoryear{Pinson, McSharry and
Madsen}{2010}]{Pinson10b}
%
\begin{barticle}[auto:STB|2013/09/19|12:14:10]
\bauthor{\bsnm{Pinson},~\bfnm{P.}\binits{P.}},
\bauthor{\bsnm{McSharry},~\bfnm{P.~E.}\binits{P.~E.}} \AND
\bauthor{\bsnm{Madsen},~\bfnm{H.}\binits{H.}}
(\byear{2010}).
\btitle{Reliability diagrams for nonparametric density forecasts of continuous
variables: Accounting for serial correlation}.
\bjournal{Quart. J. Royal Met. Soc.}
\bvolume{136}
\bpages{77--90}.
\bptok{imsref}%
\end{barticle}
%
\endbibitem

\bibitem[\protect\citeauthoryear{Pinson et~al.}{2009}]{Pinson09}
%
\begin{barticle}[auto:STB|2013/09/19|12:14:10]
\bauthor{\bsnm{Pinson},~\bfnm{P.}\binits{P.}},
\bauthor{\bsnm{Nielsen},~\bfnm{H.~Aa.}\binits{H.~A.}},
\bauthor{\bsnm{Madsen},~\bfnm{H.}\binits{H.}},
\bauthor{\bsnm{Papaefthymiou},~\bfnm{G.}\binits{G.}} \AND
\bauthor{\bsnm{Kl{\"o}ckl},~\bfnm{B.}\binits{B.}}
(\byear{2009}).
\btitle{From probabilistic forecasts to statistical scenarios of short-term
wind power production}.
\bjournal{Wind Energy}
\bvolume{12}
\bpages{51--62}.
\bptok{imsref}%
\end{barticle}
%
\endbibitem

\bibitem[\protect\citeauthoryear{Raiffa and Schaifer}{1961}]{Raiffa61}
%
\begin{bmisc}[auto:STB|2013/09/19|12:14:10]
\bauthor{\bsnm{Raiffa},~\bfnm{H.}\binits{H.}} \AND
\bauthor{\bsnm{Schaifer},~\bfnm{R.}\binits{R.}}
(\byear{1961}).
\bhowpublished{\textit{Applied Statistical Decision Theory}.
Division of Research,
Harvard Business School, Boston}.
\bptok{imsref}%
\end{bmisc}
%
\endbibitem

\bibitem[\protect\citeauthoryear{Reikard}{2008}]{Reikard08}
%
\begin{barticle}[auto:STB|2013/09/19|12:14:10]
\bauthor{\bsnm{Reikard},~\bfnm{G.}\binits{G.}}
(\byear{2008}).
\btitle{Using temperature and state transitions to forecast wind speed}.
\bjournal{Wind Energy}
\bvolume{11}
\bpages{431--443}.
\bptok{imsref}%
\end{barticle}
%
\endbibitem

\bibitem[\protect\citeauthoryear{Roulston et~al.}{2003}]{Roulston03}
%
\begin{barticle}[auto:STB|2013/09/19|12:14:10]
\bauthor{\bsnm{Roulston},~\bfnm{M.~S.}\binits{M.~S.}},
\bauthor{\bsnm{Kaplan},~\bfnm{D.~T.}\binits{D.~T.}},
\bauthor{\bsnm{Hardenberg},~\bfnm{J.}\binits{J.}} \AND
\bauthor{\bsnm{Smith},~\bfnm{L.~A.}\binits{L.~A.}}
(\byear{2003}).
\btitle{Using medium-range weather forecasts to improve the value of
wind energy
production}.
\bjournal{Renew. Energ.}
\bvolume{28}
\bpages{585--602}.
\bptok{imsref}%
\end{barticle}
%
\endbibitem

\bibitem[\protect\citeauthoryear{Sideratos and
Hatziargyriou}{2007}]{Sideratos07}
%
\begin{barticle}[auto:STB|2013/09/19|12:14:10]
\bauthor{\bsnm{Sideratos},~\bfnm{G.}\binits{G.}} \AND
\bauthor{\bsnm{Hatziargyriou},~\bfnm{N.~D.}\binits{N.~D.}}
(\byear{2007}).
\btitle{An advanced statistical method for wind power forecasting}.
\bjournal{IEEE T. Power Syst.}
\bvolume{22}
\bpages{258--265}.
\bptok{imsref}%
\end{barticle}
%
\endbibitem

\bibitem[\protect\citeauthoryear{Sideratos and
Hatziargyriou}{2012}]{Sideratos12}
%
\begin{barticle}[auto:STB|2013/09/19|12:14:10]
\bauthor{\bsnm{Sideratos},~\bfnm{G.}\binits{G.}} \AND
\bauthor{\bsnm{Hatziargyriou},~\bfnm{N.~D.}\binits{N.~D.}}
(\byear{2012}).
\btitle{Probabilistic wind power forecasting using radial basis
function neural
networks}.
\bjournal{IEEE T. Power Syst.}
\bvolume{27}
\bpages{1788--1796}.
\bptok{imsref}%
\end{barticle}
%
\endbibitem

\bibitem[\protect\citeauthoryear{Skytte}{1999}]{Skytte99}
%
\begin{barticle}[auto:STB|2013/09/19|12:14:10]
\bauthor{\bsnm{Skytte},~\bfnm{K.}\binits{K.}}
(\byear{1999}).
\btitle{The regulating power market on the Nordic power exchange Nord
Pool: An
econometric analysis}.
\bjournal{Energ. Econ.}
\bvolume{21}
\bpages{295--308}.
\bptok{imsref}%
\end{barticle}
%
\endbibitem

\bibitem[\protect\citeauthoryear{Stroud, M{\"u}ller and
Sans{\'o}}{2001}]{Stroud99}
%
\begin{barticle}[mr]
\bauthor{\bsnm{Stroud},~\bfnm{Jonathan~R.}\binits{J.~R.}},
\bauthor{\bsnm{M{\"u}ller},~\bfnm{Peter}\binits{P.}} \AND
\bauthor{\bsnm{Sans{\'o}},~\bfnm{Bruno}\binits{B.}}
(\byear{2001}).
\btitle{Dynamic models for spatiotemporal data}.
\bjournal{J. R. Stat. Soc. Ser. B Stat. Methodol.}
\bvolume{63}
\bpages{673--689}.
\bid{doi={10.1111/1467-9868.00305}, issn={1369-7412}, mr={1872059}}
\bptnote{check year}%
\bptok{imsref}%
\end{barticle}
%
\endbibitem

\bibitem[\protect\citeauthoryear{Taylor, McSharry and
Buizza}{1999}]{Taylor09}
%
\begin{barticle}[auto:STB|2013/09/19|12:14:10]
\bauthor{\bsnm{Taylor},~\bfnm{J.~W.}\binits{J.~W.}},
\bauthor{\bsnm{McSharry},~\bfnm{P.~E.}\binits{P.~E.}} \AND
\bauthor{\bsnm{Buizza},~\bfnm{R.}\binits{R.}}
(\byear{1999}).
\btitle{Wind power density forecasting using ensemble predictions and time
series models}.
\bjournal{IEEE T. Energy Conver.}
\bvolume{24}
\bpages{775--782}.
\bptok{imsref}%
\end{barticle}
%
\endbibitem

\bibitem[\protect\citeauthoryear{Tol}{1997}]{Tol97}
%
\begin{barticle}[auto:STB|2013/09/19|12:14:10]
\bauthor{\bsnm{Tol},~\bfnm{R.~S.~J.}\binits{R.~S.~J.}}
(\byear{1997}).
\btitle{Autoregressive conditional heteroscedasticity in daily wind speed
measurements}.
\bjournal{Theo. Appl. Clim.}
\bvolume{56}
\bpages{113--122}.
\bptok{imsref}%
\end{barticle}
%
\endbibitem

\bibitem[\protect\citeauthoryear{Tong}{2011}]{Tong11}
%
\begin{barticle}[mr]
\bauthor{\bsnm{Tong},~\bfnm{Howell}\binits{H.}}
(\byear{2011}).
\btitle{Threshold models in time series analysis---30 years on}.
\bjournal{Stat. Interface}
\bvolume{4}
\bpages{107--118}.
\bid{doi={10.4310/SII.2011.v4.n2.a1}, issn={1938-7989}, mr={2812802}}
\bptnote{check related}%
\bptok{imsref}%
\end{barticle}
%
\endbibitem

\bibitem[\protect\citeauthoryear{Vautard et~al.}{2010}]{Vautard10}
%
\begin{barticle}[auto:STB|2013/09/19|12:14:10]
\bauthor{\bsnm{Vautard},~\bfnm{R.}\binits{R.}},
\bauthor{\bsnm{Cattiaux},~\bfnm{J.}\binits{J.}},
\bauthor{\bsnm{Yiou},~\bfnm{P.}\binits{P.}},
\bauthor{\bsnm{Th{\'e}paut},~\bfnm{J.~N.}\binits{J.~N.}} \AND
\bauthor{\bsnm{Ciais},~\bfnm{P.}\binits{P.}}
(\byear{2010}).
\btitle{Northern Hemisphere atmospheric stilling partly attributed to an
increase in surface roughness}.
\bjournal{Nat. Geosci.}
\bvolume{3}
\bpages{756--761}.
\bptok{imsref}%
\end{barticle}
%
\endbibitem

\bibitem[\protect\citeauthoryear{Vincent et~al.}{2010}]{Vincent10}
%
\begin{barticle}[auto:STB|2013/09/19|12:14:10]
\bauthor{\bsnm{Vincent},~\bfnm{C.~L.}\binits{C.~L.}},
\bauthor{\bsnm{Giebel},~\bfnm{G.}\binits{G.}},
\bauthor{\bsnm{Pinson},~\bfnm{P.}\binits{P.}} \AND
\bauthor{\bsnm{Madsen},~\bfnm{H.}\binits{H.}}
(\byear{2010}).
\btitle{Resolving nonstationary spectral information in wind speed
time series
using the Hilbert--Huang transform}.
\bjournal{J. Appl. Meteor. Clim.}
\bvolume{49}
\bpages{253--267}.
\bptok{imsref}%
\end{barticle}
%
\endbibitem

\bibitem[\protect\citeauthoryear{Wallis}{2003}]{Wallis03}
%
\begin{barticle}[auto:STB|2013/09/19|12:14:10]
\bauthor{\bsnm{Wallis},~\bfnm{K.~F.}\binits{K.~F.}}
(\byear{2003}).
\btitle{Chi-squared tests of interval and density forecasts, and the
bank of
England's fan charts}.
\bjournal{Int. J. Forecasting}
\bvolume{19}
\bpages{165--175}.
\bptok{imsref}%
\end{barticle}
%
\endbibitem

\bibitem[\protect\citeauthoryear{Weron}{2006}]{Weron06}
%
\begin{bbook}[auto:STB|2013/09/19|12:14:10]
\bauthor{\bsnm{Weron},~\bfnm{R.}\binits{R.}}
(\byear{2006}).
\btitle{Modeling and Forecasting Electricity Loads and Prices: A Statistical
Approach}.
\bpublisher{Wiley}, \blocation{New York}.
\bptok{imsref}%
\end{bbook}
%
\endbibitem

\bibitem[\protect\citeauthoryear{Wolak}{2013}]{Wolak13}
%
\begin{bmisc}[auto:STB|2013/09/19|12:14:10]
\bauthor{\bsnm{Wolak},~\bfnm{F.~A.}\binits{F.~A.}}
(\byear{2013}).
\bhowpublished{Regulating competition in wholesale electricity supply. In
\textit{Economic Regulation and Its Reform: What Have We Learned}? (N. L.
Rose, ed.). Univ. Chicago Press. To appear}.
\bptok{imsref}%
\end{bmisc}
%
\endbibitem

\bibitem[\protect\citeauthoryear{WWEA (World Wind Energy
Association)}{2012}]{WWEA12}
%
\begin{bmisc}[auto:STB|2013/09/19|12:14:10]
\borganization{WWEA (World Wind Energy Association)}
(\byear{2012}).
\bhowpublished{2012---Half-year report. Technical report. Available at
\href{http://www.wwindea.org/webimages/Half-year\_report\_2012.pdf}{http://}
\href{http://www.wwindea.org/webimages/Half-year\_report\_2012.pdf}{www.wwindea.org/webimages/Half-year\_report\_2012.pdf}}.
\bptok{imsref}%
\end{bmisc}
%
\endbibitem

\bibitem[\protect\citeauthoryear{Zhu and Genton}{2012}]{Zhu12}
%
\begin{barticle}[mr]
\bauthor{\bsnm{Zhu},~\bfnm{Xinxin}\binits{X.}} \AND
\bauthor{\bsnm{Genton},~\bfnm{Marc~G.}\binits{M.~G.}}
(\byear{2012}).
\btitle{Short-term wind speed forecasting for power system operations}.
\bjournal{Int. Stat. Rev.}
\bvolume{80}
\bpages{2--23}.
\bid{doi={10.1111/j.1751-5823.2011.00168.x}, issn={0306-7734}, mr={2990340}}
\bptok{imsref}%
\end{barticle}
%
\endbibitem

\bibitem[\protect\citeauthoryear{Zugno, J{\'o}nsson and
Pinson}{2013}]{Zugno13b}
%
\begin{bmisc}[auto:STB|2013/09/19|12:14:10]
\bauthor{\bsnm{Zugno},~\bfnm{M.}\binits{M.}},
\bauthor{\bsnm{J{\'o}nsson},~\bfnm{T.}\binits{T.}} \AND
\bauthor{\bsnm{Pinson},~\bfnm{P.}\binits{P.}}
(\byear{2013}).
\bhowpublished{Trading wind energy based on probabilistic forecasts of both
wind generation and market quantities. \textit{Wind Energy}. \textbf{16} 909--926}.
\bptok{imsref}%
\end{bmisc}
%
\endbibitem

\bibitem[\protect\citeauthoryear{Zugno et~al.}{2013}]{Zugno13}
%
\begin{bmisc}[auto:STB|2013/09/19|12:14:10]
\bauthor{\bsnm{Zugno},~\bfnm{M.}\binits{M.}},
\bauthor{\bsnm{Morales},~\bfnm{J.~M.}\binits{J.~M.}},
\bauthor{\bsnm{Pinson},~\bfnm{P.}\binits{P.}} \AND
\bauthor{\bsnm{Madsen},~\bfnm{H.}\binits{H.}}
(\byear{2013}).
\bhowpublished{Pool strategy of a price-maker wind power producer.
\textit{IEEE
T. Power Syst.} \textbf{28} 3440--3450}.
\bptok{imsref}%
\end{bmisc}
%
\endbibitem

\end{thebibliography}
\end{document}